\documentclass[twocolumn,showpacs,preprintnumbers,amsmath,amssymb]{revtex4}
\usepackage{graphicx}
\usepackage{dcolumn}
\usepackage{bm}
\usepackage{enumerate}
\DeclareFontFamily{U}{rsfs}{\skewchar\font"7F}
\DeclareFontShape{U}{rsfs}{m}{n}{
	<-6> rsfs5
	<6-8> rsfs7
	<8-> rsfs10
	}{}
\DeclareMathAlphabet{\mathscr}{U}{rsfs}{m}{n}

\def\ii{{\mathrm{i}}}

\def\dd{{\mathrm{d}}}

\def\bracket#1{\langle #1 \rangle}

\def\sub#1{_{\mathrm{#1}}}
\def\up#1{^{\mathrm{#1}}}
\def\Vec#1{\mbox{\boldmath $#1$}}

\def\He4{\mbox{$^4$He}}
\begin{document}

\title{Kolmogorov Spectrum of Quantum Turbulence}

\author{Michikazu Kobayashi and Makoto Tsubota}

\affiliation{Faculty of Science, Osaka City University, 3-3-138, Sumiyoshi-ku, Osaka 558-8585, Japan}

\begin{abstract}
There is a growing interest in the relation between classical turbulence and quantum turbulence. Classical turbulence arises from complicated dynamics of eddies in a classical fluid. In contrast, quantum turbulence consists of a tangle of stable topological defects called quantized vortices, and thus quantum turbulence provides a simpler prototype of turbulence than classical turbulence. In this paper, we investigate the dynamics and statistics of quantized vortices in quantum turbulence by numerically solving a modified Gross-Pitaevskii equation. First, to make decaying turbulence, we introduce a dissipation term that works only at scales below the healing length. Second, to obtain steady turbulence through the balance between injection and decay, we add energy injection at large scales. The energy spectrum is quantitatively consistent with the Kolmogorov law in both decaying and steady turbulence. Consequently, this is the first study that confirms the inertial range of quantum turbulence.
\end{abstract}


\maketitle

\section{Introduction}\label{sec-introduction}

The study of turbulence has very long history, going back at least to Leonard da Vinci, and the turbulent state in fluid dynamics is now intensely studied in various fields including mathematics, and the experimental, theoretical, and computational branches of physics \cite{Batchelor, Frisch}. Until recently, only classical turbulence (CT) has been studied. Classical incompressible Newtonian fluids are usually described using the Navier-Stokes equation
\begin{subequations} \begin{align} &\frac{\partial}{\partial t}\Vec{v}(\Vec{x},t)+\Vec{v}(\Vec{x},t)\cdot\nabla\Vec{v}(\Vec{x},t)\nonumber\\&\quad =-\frac{1}{\rho}\nabla P(\Vec{x},t)+\nu\nabla^2\Vec{v}(\Vec{x},t)\label{eq-Navier-Stokes}\\ &\nabla\cdot\Vec{v}(\Vec{x},t)=0,\label{eq-continuity} \end{align} \end{subequations}
where $\Vec{v}(\Vec{x},t)$ is the velocity of fluid, $P(\Vec{x},t)$ is the external pressure, $\rho$ is the density of fluid and $\nu$ is the kinematic viscosity. Time development of $\Vec{v}(\Vec{x},t)$ is dominated by eq. (\ref{eq-Navier-Stokes}), while eq. (\ref{eq-continuity}) confirms incompressibility of fluid and determines $P(\Vec{x},t)$. Flow of this fluid can be characterized by the ratio of the second term of the left-hand side of eq. (\ref{eq-Navier-Stokes}), hereafter the inertial term, to the second term of the right-hand side, hereafter the viscous term. This ratio is the Reynolds number
\begin{equation} R=\frac{\bar{v}L}{\nu}.\label{eq-Classical-Reynolds} \end{equation}
Here $\bar{v}$ and $L$ are the characteristic velocity of flow and the characteristic scale, respectively. When $\bar{v}$ increases to the point that the Reynolds number exceeds a critical value, the system changes from a laminar to a turbulent state in which the flow is highly complicated with many eddies. These eddies have a high concentration of vorticity $\Vec{\omega}(\Vec{x},t)=\nabla\times\Vec{v}(\Vec{x},t)$. Owing to its complexity, the dynamics of the turbulent state is very difficult to investigate. Nevertheless, G. I. Taylor, a pioneer of modern studies of turbulence, built a theoretical and experimental framework for turbulence by introducing the idea in which turbulence is {\it homogeneous} and {\it isotropic} at smaller scales than the system size $L$ \cite{Taylor}. Based on this idea, Kolmogorov proposed a self-similar, equilibrium state of fully developed turbulence together with a statistical law of turbulence called the Kolmogorov law \cite{Kolmogorov-a, Kolmogorov-b}. The Kolmogorov law was confirmed through many experimental and theoretical studies, which led to an explosion in our understanding of CT. Today, numerical simulation of the Navier-Stokes equation is an indispensable tool for understanding CT because it is almost impossible to study the equation analytically. The rapid development of computers with the resulting improvements of large-scale simulations has accelerated our understanding of CT. However, two major problems have remained. One is that a large system size $L$ and a sufficient spatial resolution are required for analyzing well-developed CT at high Reynolds numbers $R$; a problem that needs higher-powered computers for its solution. The other problem is the identification of eddies in CT: there is no global way to identify classical eddies. There are several variables to identify eddies, such as the pressure, the velocity field, and the vorticity \cite{Tanaka}, but the identification of eddies in CT depends on the choice of these variables and their thresholds of identification. As a result, the dynamics of eddies, for example, the cascade process of eddies in CT, becomes only conceptual. Consequently, the existence of the cascade process, the self-similarity of the dynamics of eddies, and how they are related to the statistics of CT are not well understood.

Recently, in the field of low temperature physics, a similarity has been found between CT and turbulence in superfluid \He4. As a result, there has been much interest in studying superfluid turbulence (ST) as a way to better understand CT.

The physics of ST in liquid \He4 is one of the most important topics in low temperature physics \cite{Donnelly}. Liquid \He4 enters the superfluid state at 2.17 K. Below this temperature, the hydrodynamics is usually described using the two-fluid model in which the system consists of an inviscid superfluid and a viscous normal fluid. Early experimental studies on the subject focused on thermal counterflow in which the normal fluid and superfluid flow in opposite directions. Their flow is driven by the injected heat current, and it was found that the superflow becomes dissipative when the relative velocity between two fluids exceeds a critical value \cite{Donnelly}. Feynman proposed that this is a superfluid turbulent state consisting of a tangle of quantized vortices \cite{Feynman}. Vinen later confirmed Feynman's proposition experimentally by showing that the dissipation comes from the mutual friction between vortices and the normal flow \cite{Vinen-1}. After that, many experimental studies were done on ST in thermal counterflow systems \cite{Tough}. In particular, Schwarz clarified the picture of ST having tangled vortices by doing a large-scale numerical simulation of the quantized vortex-filament model in thermal counterflow \cite{Schwarz}. However, as thermal counterflow has no analogy with conventional fluid dynamics, it has not helped us understood the relation between ST and CT.

More recently, experiments have been done on ST that do not involve thermal counterflow. Such studies opened a new stage in studies of ST. Maurer {\it et al.} studied a ST that was driven by two counter-rotating disks \cite{Maurer} in superfluid \He4 at temperatures above 1.4 K. They measured the pressure and converted it to the local velocity fluctuation, and found that the energy spectrum was consistent with the Kolmogorov law. Stalp {\it et al.} measured the attenuation of second sound \cite{Stalp} in superfluid \He4 at temperatures above 1 K, and found that inferred decay of grid turbulence was consistent with the Kolmogorov law. These experiments are consistent in the sense that they show similarities between ST and CT. This similarity can be understood using the idea that the superfluid and the normal fluid are likely to be coupled together by the mutual friction between them and thus to behave like a conventional fluid \cite{Vinen-2}. However, the normal fluid is negligible at very low temperatures, and this brings up an important question: even without the normal fluid, is ST still similar to CT? Such a fluid is a pure quantum fluid and its turbulence can be called {\it quantum turbulence}, hereafter QT.

To address this question, we consider the statistical law of CT \cite{Frisch}. For fully developed turbulence of an incompressible classical fluid, the energy is injected into the fluid at scales comparable to the system size $L$ in the energy-containing range, and in the inertial range, this energy is transferred to smaller scales without being dissipated. In this range, the system is locally homogeneous and isotropic, which leads to the statistics of the energy spectrum known as the Kolmogorov law
\begin{equation} E(k)=C\varepsilon^{2/3}k^{-5/3}.\label{eq-Kolmogorov} \end{equation}
Here the energy spectrum $E(k)$ is defined as $E=\int\dd k\:E(k)$, where $E$ is the kinetic energy per unit mass and $k$ is the wave number from the Fourier transformation of the velocity field. The energy transferred to smaller scales in the energy-dissipative range is dissipated through the viscosity of the fluid with the dissipation rate $\varepsilon$ in eq. (\ref{eq-Kolmogorov}), which is equal to the energy flux $\Pi$ in the inertial range. The Kolmogorov constant $C$ is a dimensionless parameter of order unity \cite{Kida}.

The inertial range is thought to be sustained by the self-similar Richardson cascade in which large eddies are broken up into smaller ones with many reconnections. In CT, however, the Richardson cascade is not completely understood because it is impossible to definitely identify each eddy - the eddies keep nucleating and annihilating due to the viscosity. In contrast, quantized vortices in a quantum fluid are definite and stable topological defects. At low temperatures, a macroscopic number of particles in a Bose system, such as liquid \He4, occupy a single particle ground state. In this case, matter waves of Bose particles form a coherent scalar macroscopic wave function $\Phi(\Vec{x},t)=f(\Vec{x},t)\exp[\ii\phi(\Vec{x},t)]$, which is known as Bose-Einstein condensation. The dynamics of the macroscopic wave function is governed by the Gross-Pitaevskii (GP) equation \cite{Gross,Pitaevskii}
\begin{equation} \ii\frac{\partial}{\partial t}\Phi(\Vec{x},t)=[-\nabla^2-\mu+g|\Phi(\Vec{x},t)|^2]\Phi(\Vec{x},t).\label{eq-GP} \end{equation}
The GP equation is obtained from the Hamilton-Jacobi equation for the Hamiltonian
\begin{equation} H=\int\dd\Vec{x}\Big[|\nabla\Phi(\Vec{x},t)|^2-\mu|\Phi(\Vec{x},t)|^2+\frac{g}{2}|\Phi(\Vec{x},t)|^4\Big],\label{eq-Hamiltonian} \end{equation}
where $\mu$ is the chemical potential, $g$ is the coupling constant, $|\Phi(\Vec{x},t)|^2=f(\Vec{x},t)^2$ is the condensate density, and the superfluid velocity $\Vec{v}(\Vec{x},t)$ is given by $\Vec{v}(\Vec{x},t)=2\nabla\phi(\Vec{x},t)$. The vorticity $\Vec{\omega}(\Vec{x},t)=\nabla\times\Vec{v}(\Vec{x},t)$ vanishes everywhere in a single-connected region of the fluid and thus, all rotational flow is carried by quantized vortices. In the core of each vortex, $\Phi(\Vec{x},t)$ vanishes so that the circulation $\oint\Vec{v}\cdot \Vec{\dd s}$ around the core is quantized by $4\pi$. The vortex core size is given by the healing length $\xi=1/f\sqrt{g}$. Quantized vortices at finite temperatures can decay through mutual friction with the normal fluid at any scale. At very low temperatures the vortices can decay by 1) the emission of compressible excitation through vortex reconnections \cite{Leadbeater}, and 2) a process in which vortices reduce through the Richardson cascade process and eventually change to compressible excitations. These are the only two decay mechanisms at very low temperature. In any case, dissipation occurs only at scales below the healing length; for larger scales, we can obtain the turbulent state at high Reynolds numbers that is free from dissipation by quantized vortices. Therefore, QT at very low temperatures gives the high Reynolds number and the real Richardson cascade of definite quantized vortices, and thus can become an ideal prototype to study statistics such as the Kolmogorov law and the Richardson cascade in the inertial range of CT.

There are two formulations for studying the dynamics of quantized vortices in QT. One is the vortex-filament model \cite{Schwarz}, and the other is the GP model. By using the vortex-filament model with no normal fluid component, Araki {\it et al.} studied numerically a vortex tangle arising from Taylor-Green flow, and obtained an energy spectrum consistent with the Kolmogorov law \cite{Araki}. By eliminating the smallest vortices, which had a size comparable to the numerical space resolution, they introduced dissipation into the system and realized decaying turbulence. Nore {\it et al.} used the GP equation to numerically study the energy spectrum of QT \cite{Nore}. They showed that the kinetic energy consists of a compressible part due to compressible excitations and an incompressible part coming from quantized vortices. Compressible excitations of wavelength smaller than the healing length are created through vortex reconnections or through the disappearance of small vortex loops \cite{Leadbeater,Ogawa}, so that the incompressible kinetic energy changes to compressible kinetic energy while conserving the total energy. The spectrum of the incompressible kinetic energy is consistent with the Kolmogorov law in a short period \cite{Cichowlas}. However, the consistency is broken in the late stage when many compressible excitations are created \cite{Leadbeater, Ogawa, Berloff-1, Berloff-2} and hinder the Richardson cascade process, even on the large scales.

Our approach here is to introduce a dissipation term that works only on scales smaller than the healing length $\xi$. This dissipation term can account for dissipation processes such as the scattering between the particle reservoir and the emitted compressible excitations that can occur only at scales smaller than $\xi$ even at $T=0$ K \cite{Leadbeater,Ogawa}. This dissipation removes only short-wavelength excitations (not vortices), thus preventing the vortices from receiving the excitation energy. Compared to the usual GP model, this approach enables us to more clearly study the relation between QT and CT. The Richardson cascade process at the large scales must be independent of the details of the dissipative mechanism at the small scales.

We use our modified GP equation to investigate two types of QT. For decaying turbulence, we start from random configurations of the phase of the wave function and study the decay. We provide here a more detailed study of this type than that discussed in our previous paper \cite{Kobayashi}. For steady turbulence (the other type), we assume that energy is injected into the system at a large scale by a moving random potential. For both types, compressible excitations are highly suppressed compared to quantized vortices, and the spectrum of the incompressible kinetic energy is found to be consistent with the Kolmogorov law, which verifies that QT is similar to CT. We also show here that steady turbulence with the moving random potential clarifies the energy-containing range, the inertial range and the energy-dissipative range. By calculating the energy dissipation rate and the energy flux in the wave number space, we confirmed the following scenario. First, vortices are nucleated by the moving random potential in the energy-containing range. Second, the vortices decay into smaller vortices in the inertial range through the Richardson cascade process. Finally, the smallest vortices change to compressible excitations, which are then removed by the introduced dissipation. Although this scenario of quantized vortices in QT is very similar to that believed about eddies in CT, it is much clearer, because quantized vortices are definite topological defects.

This paper is organized as follows. In Sec. \ref{sec-model}, we review our model of the modified GP equation that includes small-scale dissipation. After discussing the numerical procedure of our model in Sec. \ref{sec-procedure}, we show numerical simulations of i) a straight vortex line in Sec. \ref{sec-single-vortex-line}, ii) decaying turbulence in Sec. \ref{sec-decay}, and iii) steady turbulence in Sec. \ref{sec-steady}. Section \ref{sec-conclusion} is devoted to our conclusions.

\section{GP Equation with Small-Scale Dissipation}\label{sec-model}

In this section, we discuss the dynamics of the original GP equation (\ref{eq-GP}) with quantized vortices and why the compressible short-wavelength excitations make it difficult to study QT. Then we propose a modified GP equation with small-scale dissipation. The modified GP equation gives a clear criterion for the inertial range of QT and has a form similar to that of the Navier-Stokes equation with large effective Reynolds number.

\begin{figure*}[htb] \centering \begin{minipage}[t]{0.24\linewidth} \begin{center} \includegraphics[width=.99\linewidth]{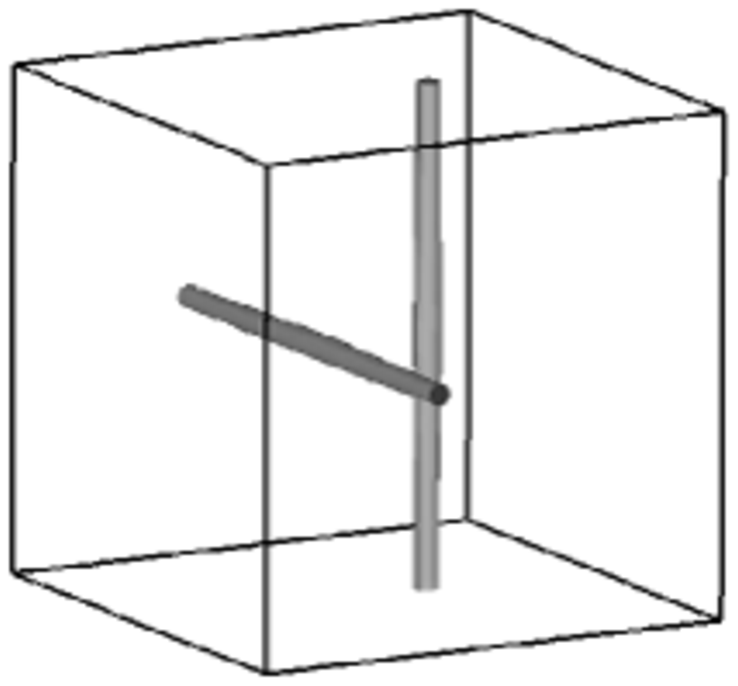}\\ (a) \end{center} \end{minipage} \begin{minipage}[t]{0.24\linewidth} \begin{center} \includegraphics[width=.99\linewidth]{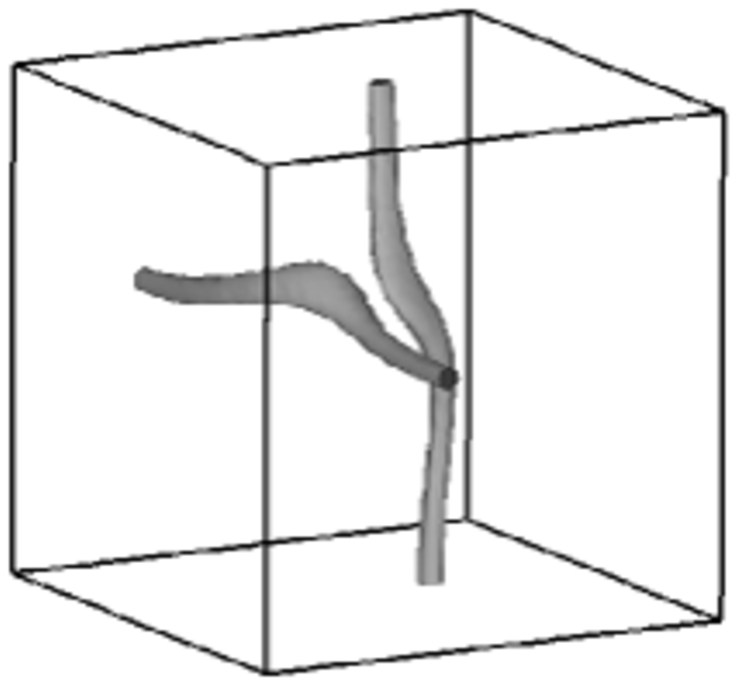}\\ (b) \end{center} \end{minipage}\begin{minipage}[t]{0.24\linewidth} \begin{center} \includegraphics[width=.99\linewidth]{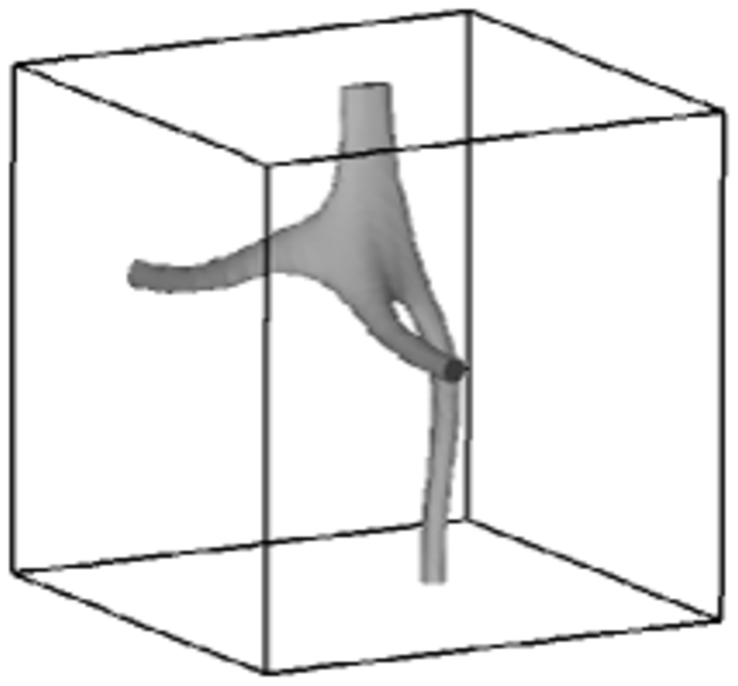}\\ (c) \end{center} \end{minipage}\begin{minipage}[t]{0.24\linewidth} \begin{center} \includegraphics[width=.99\linewidth]{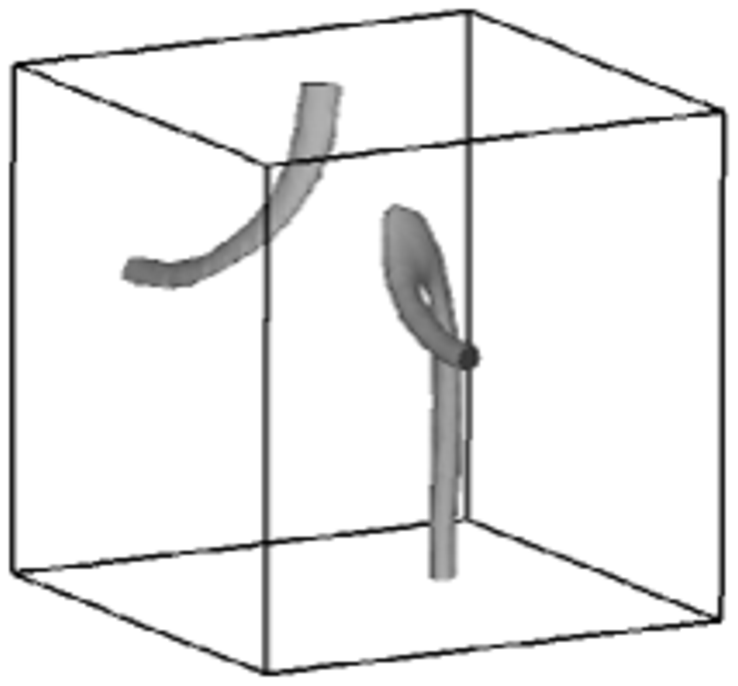}\\ (d) \end{center} \end{minipage}
\caption{\label{fig-reconnection} The reconnection of two quantized vortices starting from a skew position of two straight vortex lines. (a) Initial state. (b) Just before the two vortices connect. (c) Connection of two vortex lines. (d) After the vortex lines separate in a newly connected arrangement, also called reconnection. Contours in all figures are high concentrations of vorticity (98\% of maximum vorticity).} \end{figure*}

The dynamics of quantized vortices was analyzed by Koplik {\it et al} \cite{Koplik}. and Ogawa {\it et al}. \cite {Ogawa} by using the GP equation (\ref{eq-GP}). Through the Madelung transformation $\Phi(\Vec{x},t)=f(\Vec{x},t)\exp[\ii\phi(\Vec{x},t)]$, the GP equation (\ref{eq-GP}) becomes
\begin{align} f(\Vec{x},t)\frac{\partial}{\partial t}\phi(\Vec{x},t)=&\nabla^2f(\Vec{x},t)-f(\Vec{x},t)\nabla\phi(\Vec{x},t)^2\nonumber\\ &-[gf(\Vec{x},t)^2-\mu]f(\Vec{x},t).\label{eq-initial-Madelung-GP} \end{align}
At scales larger than $\xi$, we can neglect the curvature of the amplitude of the wave function near quantized vortices $\nabla f(\Vec{x},t)$. With this assumption, the equation for the vorticity $\omega(\Vec{x},t)=\nabla\times\Vec{v}(\Vec{x},t)=2\nabla\times\nabla\phi(\Vec{x},t)$ becomes
\begin{equation} \frac{\partial}{\partial t}\Vec{\omega}(\Vec{x},t)+\frac{1}{2}\nabla\times\nabla\Vec{v}(\Vec{x},t)^2=0.\label{eq-vorticity-dynamics} \end{equation}
Thus, the dynamics of quantized vortices depends only on the fluid velocity $\Vec{v}(\Vec{x},t)$, which is similar to that of isolated eddies in a perfect classical fluid. Except for the dynamics of vortices described in eq. (\ref{eq-vorticity-dynamics}), there are two non trivial processes of quantized vortices; one is the reconnection of two vortices and the other is the disappearance of small vortex loops. Figure \ref{fig-reconnection} shows an example of vortex reconnection given by the numerical calculation of eq. (\ref{eq-GP}) starting from two straight vortex lines in a skewed orientation.
These two vortices approach and become locally anti-parallel, and then the reconnection occurs. After the reconnection, the vortices tend to be bent, which allows smaller structure to form, thus sustaining the Richardson cascade process. Ogawa {\it et al}. show that this process does not violate Kelvin's circulation theorem; moreover, the process becomes irreversible due to the emission of compressible excitations that have a wavelength that is smaller than $\xi$. In the scales, its dynamics is strongly disturbed by compressible short-wavelength excitations described by $\nabla f(\Vec{x},t)$, which is neglected in eq. (\ref{eq-vorticity-dynamics}). In turbulent states, such excitations are rapidly created through reconnections and disappearances of small vortex loops. Therefore, the dynamics of QT is affected by many compressible excitations and one cannot see the proper dynamics of quantized vortices from eq. (\ref{eq-vorticity-dynamics}) by itself.

To obtain QT free from compressible short-wavelength excitations, we have to introduce a dissipation term in the GP equation. First, we discuss a possible dissipation mechanism in the GP equation. Next, the dissipation term is arranged so that it can remove short-wavelength excitations. We assume that the system is described by $\Phi(\Vec{x},t)$ such that the energy and particles are exchanged with a particle reservoir. The particle reservoir is a thermodynamic environment that lets the chemical potential of the system equal that of the reservoir. The interaction with the particle reservoir is provided by an imaginary term in the Hamiltonian eq. (\ref{eq-Hamiltonian})
\begin{align} H=&\int\dd\Vec{x}\Big[|\nabla\Phi(\Vec{x},t)|^2-\mu|\Phi(\Vec{x},t)|^2\nonumber\\ &+\frac{g}{2}|\Phi(\Vec{x},t)|^4-\ii\Gamma(\Vec{x},t)|\Phi(\Vec{x},t)|^2\Big],\label{eq-dissipative-Hamiltonian} \end{align}
which yields the dissipative GP equation
\begin{align} &\ii\frac{\partial}{\partial t}\Phi_\mu(\Vec{x},t)\nonumber\\ &\quad =[-\nabla^2+g|\Phi_\mu(\Vec{x},t)|^2-\ii \Gamma(\Vec{x},t)]\Phi_\mu(\Vec{x},t).\label{eq-basic-GP-middle} \end{align}
Here the wave function $\Phi_\mu(\Vec{x},t)$ includes the chemical potential $\mu$ through the gauge transformation $\Phi_\mu(\Vec{x},t)=\Phi(\Vec{x},t)\exp(-\ii\mu t)$. Since $\Gamma(\Vec{x},t)$ comes from the difference of the chemical potential, we can write $\Gamma(\Vec{x},t)=\gamma(\Vec{x},t)(\mu-\mu\sub{pr})$ with the chemical potential of the particle reservoir being $\mu\sub{pr}$. We assume that the system is nearly in equilibrium with the particle reservoir and that $\Gamma(\Vec{x},t)$ is proportional to the difference in the chemical potentials. Using the approximation
\begin{equation} \ii\frac{\partial}{\partial t}\Phi_\mu(\Vec{x},t)\simeq\mu\Phi_\mu(\Vec{x},t)\label{eq-equilibrium-approximation} \end{equation}
and $\mu\simeq\mu\sub{pr}$, eq. (\ref{eq-basic-GP-middle}) becomes
\begin{align} &[\ii-\gamma(\Vec{x},t)]\frac{\partial}{\partial t}\Phi_\mu(\Vec{x},t)\nonumber\\ &\quad =[-\nabla^2+g|\Phi_\mu(\Vec{x},t)|^2+\ii\mu\gamma(\Vec{x},t)]\Phi_\mu(\Vec{x},t).\label{eq-trick-GP} \end{align}
Substituting $\Phi_\mu(\Vec{x},t)=\Phi(\Vec{x},t)\exp(-\ii\mu t)$ into eq. (\ref{eq-trick-GP}), we finally obtain the modified GP equation with the effective dissipation term $\gamma(\Vec{x},t)$ \cite{Kasamatsu}
\begin{align} &[\ii-\gamma(\Vec{x},t)]\frac{\partial}{\partial t}\Phi(\Vec{x},t)\nonumber\\ &\quad =[-\nabla^2-\mu(t)+g|\Phi(\Vec{x},t)|^2]\Phi(\Vec{x},t).\label{eq-basic-GP} \end{align}
Introduction of $\gamma(\Vec{x},t)$ conserves neither the energy nor the number of particles. However, for studying the hydrodynamics of turbulence, it is realistic to assume that the number of particles is conserved. Hence, we consider the time dependence of the chemical potential $\mu(t)$ so that the total number $N=\int\dd\Vec{x}\: |\Phi(\Vec{x},t)|^2$ can be conserved. The approximation of eq. (\ref{eq-equilibrium-approximation}) gives the following form for the chemical potential
\begin{equation} \mu(t)=\frac{1}{N}\int\dd\Vec{x}\:\Phi^\ast(\Vec{x},t)[-\nabla^2+g|\Phi(\Vec{x},t)|^2]\Phi(\Vec{x},t).\label{eq-time-chemical-potential} \end{equation}

We consider the spatial and temporal dependence of $\gamma(\Vec{x},t)$ so that this term can dissipate the short-wavelength excitations. The Fourier transformation of the GP equation (\ref{eq-basic-GP}) can be written as
\begin{align} &[\ii-\tilde{\gamma}(\Vec{k})]\frac{\partial}{\partial t}\tilde{\Phi}(\Vec{k},t)\nonumber\\ &\quad =[k^2-\mu(t)]\tilde{\Phi}(\Vec{k},t)+\frac{g}{L^6}\sum_{\Vec{k}_1,\Vec{k}_2}\tilde{\Phi}(\Vec{k}_1,t)\tilde{\Phi}^\ast(\Vec{k}_2,t)\nonumber\\ &\qquad\times\tilde{\Phi}(\Vec{k}-\Vec{k}_1+\Vec{k}_2,t).\label{eq-Fourier-GP} \end{align}
Here $\tilde{\Phi}(\Vec{k},t)$ is the spatial Fourier component of $\Phi(\Vec{x},t)$ and $L$ is the system size. $\tilde{\gamma}(\Vec{k})$ is the wave number representation of the dissipation term. For the dissipation term to remove only compressible excitations of wavelength smaller than $\xi$, it should have the form $\tilde{\gamma}(\Vec{k})=\gamma_0\theta(k-2\pi/\xi)$ with the step function $\theta$. This form is based on the idea that only compressible short-wavelength excitations interact with the particle reservoir. Because system is dissipationless at scales exceeding $\xi$, we can investigate proper vortex dynamics at large scales without considering dissipation.

Next we consider the Reynolds number of QT. By using the Madelung transformation $\Phi(\Vec{x},t)=f(\Vec{x},t)\exp[\ii\phi(\Vec{x},t)]$, eq. (\ref{eq-basic-GP}) has the hydrodynamic form in the first order of $\gamma(\Vec{x},t)$
\begin{align} &f(\Vec{x},t)\frac{\partial}{\partial t}\phi(\Vec{x},t)\nonumber\\ &\quad =\gamma(\Vec{x},t)[2\nabla f(\Vec{x},t)\cdot\nabla\phi(\Vec{x},t)+f(\Vec{x},t)\nabla^2\phi(\Vec{x},t)]\nonumber\\ &\qquad +\nabla^2f(\Vec{x},t)-f(\Vec{x},t)[\nabla\phi(\Vec{x},t)]^2\nonumber\\ &\qquad -[gf(\Vec{x},t)^2-\mu(t)]f(\Vec{x},t).\label{eq-Madelung-GP} \end{align}
By neglecting the term $\nabla f(\Vec{x},t)$ and $\nabla\gamma(\Vec{x},t)$ (because they are ineffective at scales exceeding $\xi$), we obtain the Navier-Stokes-like equation of the superfluid velocity $\Vec{v}(\Vec{x},t)=2\nabla\phi(\Vec{x},t)$,
\begin{align} &\frac{\partial}{\partial t}\Vec{v}(\Vec{x},t)+\frac{1}{2}\nabla\Vec{v}(\Vec{x},t)^2\nonumber\\ &\quad =\gamma(\Vec{x},t)\nabla[\nabla\cdot\Vec{v}(\Vec{x},t)]-2\nabla[gf(\Vec{x},t)^2-\mu(t)].\label{eq-Navier-Stokes-like} \end{align}
A comparison with the Navier-Stokes equation (\ref{eq-Navier-Stokes}) shows that the second term on the left-hand side and the first term on the right-hand side of eq. (\ref{eq-Navier-Stokes-like}) can be regarded as the inertial and dissipation terms, respectively, of the GP equation, and also that $\gamma(\Vec{x},t)$ acts as the viscosity in a conventional fluid. Following this comparison, the Reynolds number $R\sub{QT}$ of QT is
\begin{equation} R\sub{QT}=\frac{\bar{v}L}{\bar{\gamma}},\label{eq-GP-Reynolds} \end{equation}
where $\bar{v}$ is the mean velocity of QT. Here $\bar{\gamma}$ is the average of $\gamma(\Vec{x},t)$ over space and time. The average over the whole volume corresponds to the zero-wave number component of the Fourier transformation, namely $\bar{\gamma}\sim\tilde{\gamma}(\Vec{k}=0)$. However, the average over the sub-volume of dimension smaller than $\xi$ becomes $\bar{\gamma}\sim\tilde{\gamma}(|\Vec{k}|>2\pi/\xi)=\gamma_0$. Consequently, we obtain a scale-dependent Reynolds number $R\sub{QT}$ which is basically different from that of classical turbulence. At scales exceeding $\xi$, the Reynolds number $R\sub{QT}$ thus becomes infinite, which means that scales with $k<2\pi/\xi$ can be regarded as the inertial range of QT. At scales below $\xi$, $R\sub{QT}$ has a finite value $\bar{v}L/\gamma_0$.

\section{Numerical Procedure}\label{sec-procedure}

To solve the GP equation numerically with high accuracy, we use the pseudo-spectral method in space with periodic boundary conditions in a box with spatial resolution containing $N\sub{grid}=256^3$ grid points. In this box, $\Phi(\Vec{x},t)$ is discretized to $\Phi[\Vec{l}\Delta x,m\Delta t]$, $\tilde{\Phi}(\Vec{k},t)$ is discretized to $\tilde{\Phi}[\Vec{l}\Delta k,m\Delta t]$, $\mu(t)$ is discretized to $\mu[m\Delta t]$, and $\tilde{\gamma}(\Vec{k})$ is discretized to $\tilde{\gamma}[\Vec{l}\Delta k]$. The first two are related through the Fourier transformation
\begin{align} \Phi[\Vec{l}\Delta x,m\Delta t]=&\frac{1}{N\sub{grid}\Delta x^3}\sum_{\Vec{l}^\prime}\tilde{\Phi}[\Vec{l}^\prime\Delta k,m\Delta t]\nonumber\\ &\times\exp[\ii(\Vec{l}^\prime\cdot\Vec{l})\Delta k\Delta x],\label{eq-FFT} \end{align}
where integers $m$ and $\Vec{l}=[l_x,l_y,l_z]$ satisfy $m\ge 0$ and $0\le l_x,l_y,l_z\le 256$. Also, the resolutions $\Delta x$, $\Delta k=2\pi/L$ and $\Delta t$ are for space, wave number, and time. To obtain $\tilde{\Phi}[\Vec{l}\Delta k,(m+1)\Delta t]$ from $\tilde{\Phi}[\Vec{l}\Delta k,m\Delta t]$, we have to calculate the right-hand side of eq. (\ref{eq-Fourier-GP}). However, it takes a long time ($O(N\sub{grid}^3)$) to calculate the second term, so we instead calculate $h(\Vec{x},t)=g|\Phi(\Vec{x},t)|^2\Phi(\Vec{x},t)$ and its Fourier transformation $\tilde{h}(\Vec{k},t)$. Then the GP equation (\ref{eq-Fourier-GP}) becomes
\begin{equation} [\ii-\tilde{\gamma}(\Vec{k})]\frac{\partial}{\partial t}\tilde{\Phi}(\Vec{k},t)=[k^2-\mu(t)]\tilde{\Phi}(\Vec{k},t)+\tilde{h}(\Vec{k},t).\label{eq-pseudo-Fourier-GP} \end{equation}
Because $h(\Vec{x},t)$ is the functional of $\Phi(\Vec{x},t)$, we should calculate it again from $\tilde{\Phi}[\Vec{l}\Delta k,(m+1)\Delta t]$ by using the Fast-Fourier-Transformation (FFT) \cite{Press} and the inverse FFT, for the next step $\tilde{\Phi}[\Vec{l}\Delta k,(m+2)\Delta t]$. This procedure reduces the computation time from $O(N\sub{grid}^3)$ to $O(N\sub{grid}\log N\sub{grid})$ of the FFT. At each time step, the wave function is normalized, so that the total number of particles $N$ can be conserved, and the chemical potential is calculated using eq. (\ref{eq-time-chemical-potential}).

In this paper, we consider the case of $g=1$. For numerical parameters, we use a spatial resolution of $\Delta x=0.125$ and $L=32$, where the length scale is normalized by the healing length $\xi$. The wave number resolution is $\Delta k=2\pi/L\simeq 0.196$. The numerical time evolution of eq. (\ref{eq-pseudo-Fourier-GP}) was calculated using the Runge-Kutta-Verner method \cite{Press} with the time resolution $\Delta t=1\times10^{-4}$. The flowchart of the calculation is as follows:
\begin{enumerate}[(i)]
  \item Calculate $h[\Vec{l}\Delta x,m\Delta t]$ from $\Phi[\Vec{l}\Delta x,m\Delta t]$.
  \item Calculate $\tilde{h}[\Vec{l}\Delta k,m\Delta t]$ by the FFT.
  \item Calculate $\tilde{\Phi}[\Vec{l}\Delta k,(m+1)\Delta t]$ from eq. (\ref{eq-pseudo-Fourier-GP}).
  \item Calculate $\Phi[\Vec{l}\Delta x,(m+1)\Delta t]$ using the inverse FFT.
  \item $\Phi[\Vec{l}\Delta x,(m+1)\Delta t]$ is normalized to conserve $N$.
  \item Calculate $\mu[(m+1)\Delta t]$ from eq. (\ref{eq-time-chemical-potential}).
\end{enumerate}

In the framework of our numerical calculation, we cannot deal with the large wave number region $\Vec{k}>\Vec{k}\sub{c}=(2\pi/\Delta x,2\pi/\Delta x,2\pi/\Delta x)$. Throughout the FFT, the component of the wave function $\tilde{\Phi}(\Vec{k}>\Vec{k}\sub{c},t)$ with such a large wave number cycles back to that with a small wave number, a numerical process that causes an aliasing error. To reduce the aliasing error, we placed a large number of grid points $N\sub{grid}$ in a box and set the value of $\Vec{k}\sub{c}$ such that the wave function is negligible near $\Vec{k}\sub{c}$. We found that $N\sub{grid}=256^3$ is large enough to neglect the aliasing error for our QT. We discuss this problem further in section \ref{sec-decay}

\section{Dynamics of a Single Straight Vortex Line with Dissipation}\label{sec-single-vortex-line}

Before running the QT simulations, we investigated the effect of dissipation on the dynamics of a single vortex line. Here, we assume a flow, and then consider the resulting drag force due to dissipation. We use the scale-dependent dissipation term $\tilde{\gamma}(\Vec{k})=\gamma_0\theta(k-2s\pi/\xi)$ by introducing the parameter $s$, and study how the vortex dynamics depend on $s$. When $s\gtrsim 1$ and the dissipation works only on scales below $\xi$, a vortex line moves free from these forces. Therefore, the dissipation term with $s=1$ acts only on the short-wavelength excitations and does not affect the vortex dynamics at scales exceeding $\xi$. Thus $s=1$ is suitable for this study.

\begin{figure*}[htb] \centering \begin{minipage}[t]{0.24\linewidth} \begin{center} \includegraphics[width=.99\linewidth]{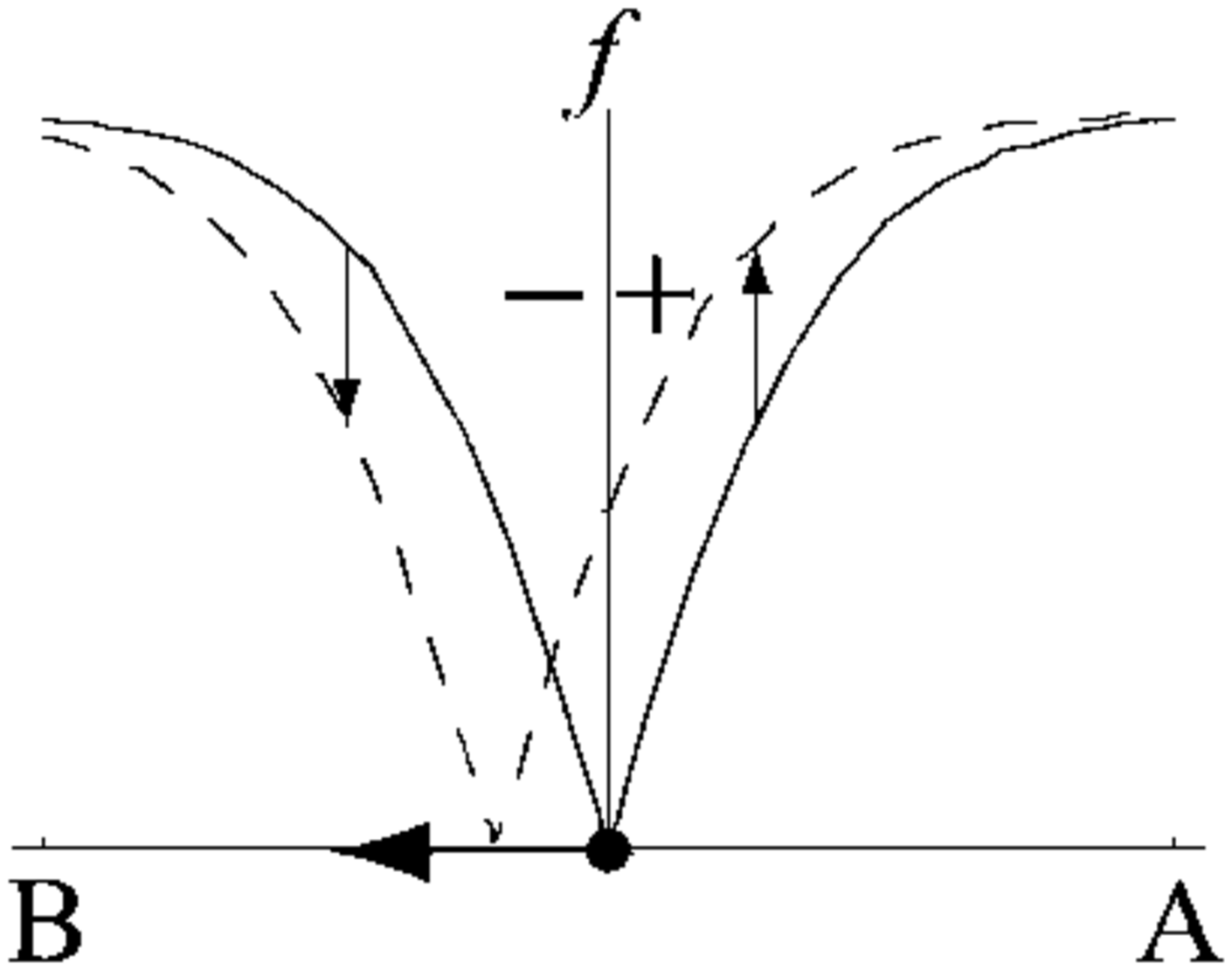}\\ (a) \end{center} \end{minipage} \begin{minipage}[t]{0.24\linewidth} \begin{center} \includegraphics[width=.99\linewidth]{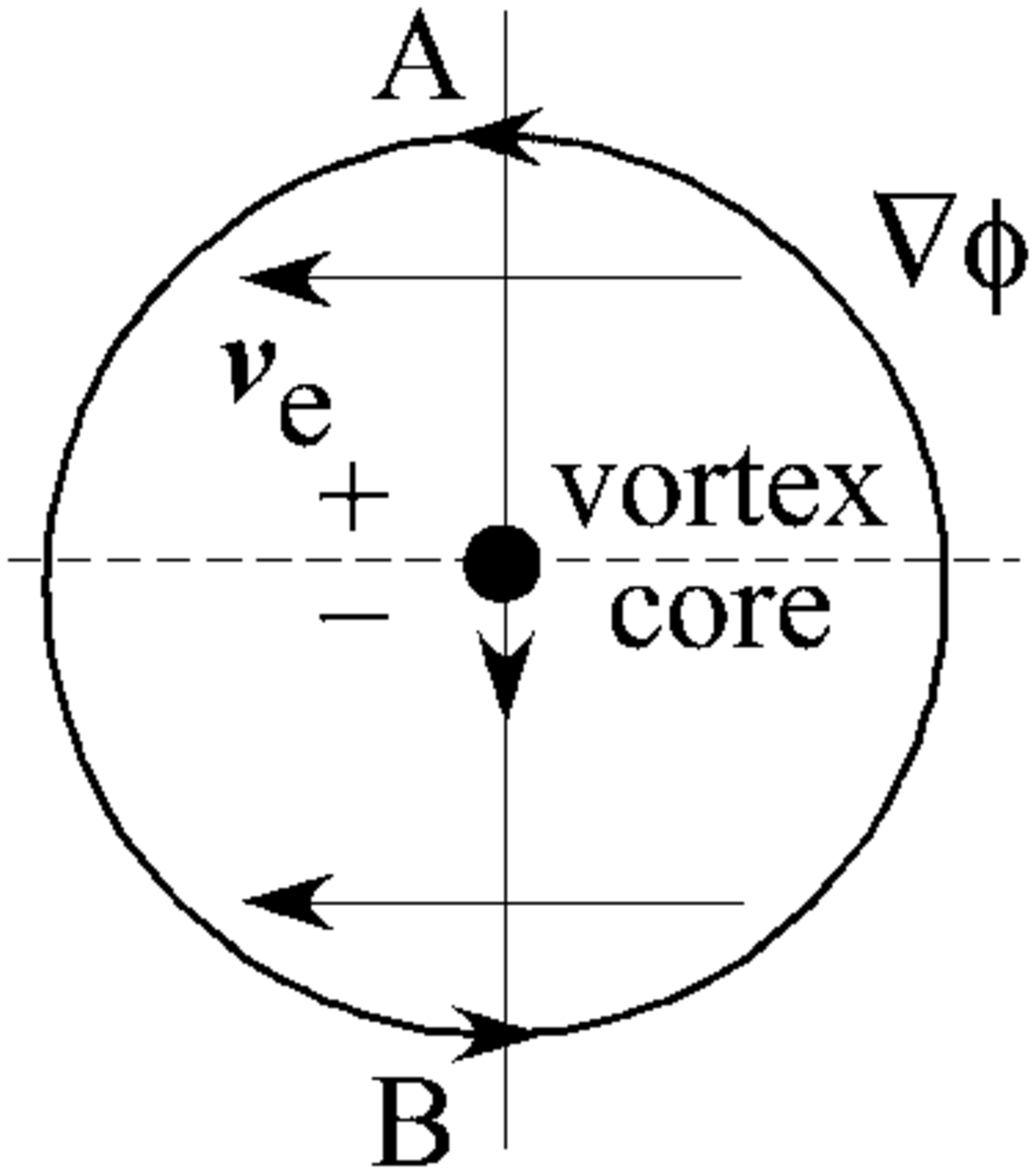}\\ (b) \end{center} \end{minipage} \begin{minipage}[t]{0.24\linewidth} \begin{center} \includegraphics[width=.99\linewidth]{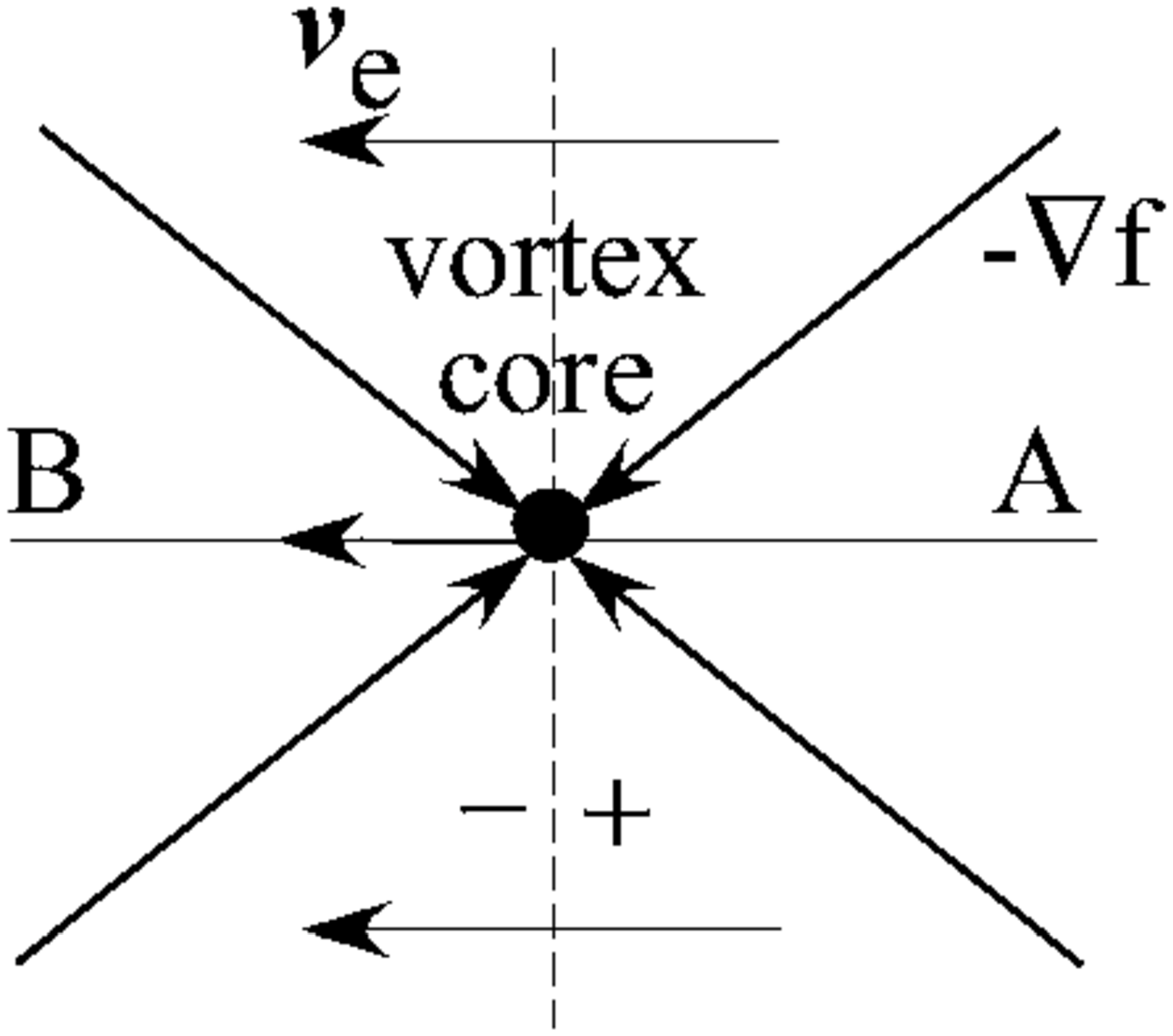}\\ (c) \end{center} \end{minipage}
\caption{\label{fig-vortex-moving} (a) The amplitude of the wave function $f(\Vec{x},t)$ at $t$ (solid line) and $t+\Delta t$ (dashed line) across the line A-B in figure (b) and (c). (b) Movement of the vortex line in a cross section of the $x-y$ plane, caused by the term $\Vec{v}\sub{e}\cdot\nabla\phi(\Vec{x},t)$ in eq. (\ref{eq-single-vortex-dynamics}). (c) Same as (b) except for the term $-\Vec{v}\sub{e}\cdot\nabla f(\Vec{x},t)$. The dot at the center is the vortex core, the thick arrows are $\nabla\phi(\Vec{x},t)$ (b) and $-\nabla f(\Vec{x},t)$ (c), thin arrows are $\Vec{v}\sub{e}$, and A and B represent the points in the positive and negative regions of $\Vec{v}\sub{e}\cdot\nabla\phi(\Vec{x},t)$ (b) and $-\Vec{v}\sub{e}\cdot\nabla f(\Vec{x},t)$ (c).} \end{figure*}

The steady solution of the GP equation with a single vortex line along the $z$ axis is represented by
\begin{equation} \Phi(\Vec{x})=f(r)\exp(\ii\varphi),\label{eq-single-vortex} \end{equation}
where $r=\sqrt{x^2+y^2}$ and $\varphi=\tan^{-1}(y/x)$ are the radius and angle in cylindrical coordinates. The amplitude of the wave function $f(r)$ follows the equation
\begin{equation} \frac{\dd^2f(r)}{\dd r^2}+\frac{1}{r}\frac{\dd f(r)}{\dd r}-\frac{f(r)}{r^2}+f(r)-f^3(r)=0,\label{eq-single-vortex-amplitude} \end{equation}
and is obtained numerically by the Shooting-Runge-Kutta-Verner method \cite{Press} with the boundary conditions $f(0)=0$ and $f(r\to\infty)=1$. Starting from the single vortex solution (\ref{eq-single-vortex}), we calculate the GP equation under a flow $\Vec{v}\sub{e}$ as
\begin{align} &[\ii-\gamma(\Vec{x},t)]\frac{\partial}{\partial t}\Phi(\Vec{x},t)\nonumber\\ &\quad =[-\nabla^2-\mu+\ii\Vec{v}\sub{e}\cdot\nabla+g|\Phi(\Vec{x},t)|^2]\Phi(\Vec{x},t).\label{eq-vorticity-GP} \end{align}
The Madelung transformation of eq. (\ref{eq-vorticity-GP}) yields
\begin{align} \frac{\partial f(\Vec{x},t)}{\partial t}=&\Vec{v}\sub{e}\cdot\nabla f(\Vec{x},t)+\gamma(\Vec{x},t) f(\Vec{x},t)\:\Vec{v}\sub{e}\cdot\nabla\phi(\Vec{x},t)\nonumber\\ &-\gamma(\Vec{x},t)^2\:\Vec{v}\sub{e}\cdot\nabla f(\Vec{x},t),\label{eq-single-vortex-dynamics} \end{align}
which is second order in $\gamma(\Vec{x},t)$. The first term on the right-hand side is the inertial term for $\Vec{v}\sub{e}$. The second and third terms of the right-hand side describe the motion of the vortex line by the dissipation term $\gamma(\Vec{x},t)$. The direction of vortex movement is indicated by the sign of $\partial f(\Vec{x},t)/\partial t$ in eq. (\ref{eq-single-vortex-dynamics}), as shown in Fig. \ref{fig-vortex-moving} (a).
As shown in this figure, the vortex moves from the side of increasing $f(\Vec{x},t)$ with time to that of decreasing $f(\Vec{x},t)$, that is, the positive to negative sides of $\partial f(\Vec{x},t)/\partial t$. Figures \ref{fig-vortex-moving} (b) and (c) show the directions of vortex motion caused by the second and third terms of the right hand side in eq. (\ref{eq-single-vortex-dynamics}) respectively. The thin arrows in these figures show the direction of $\Vec{v}\sub{e}$, thick arrows show the direction of $\nabla\phi(\Vec{x},t)$ and $-\nabla f(\Vec{x},t)$ in Fig. \ref{fig-vortex-moving} (b) and (c), respectively, and the dashed lines show the boundary between the positive and negative regions of $\Vec{v}\sub{e}\cdot\nabla\phi(\Vec{x},t)$ and $-\Vec{v}\sub{e}\cdot\nabla f(\Vec{x},t)$. For the case of Fig. \ref{fig-vortex-moving} (b), $\Vec{v}\sub{e}\cdot\nabla\phi(\Vec{x},t)$ becomes positive in upper side (and negative in the lower side), making the vortex move to the lower side. Consequently the second term in eq. (\ref{eq-single-vortex-dynamics}) causes a drag force to act in the direction perpendicular to $\Vec{v}\sub{e}$. By using similar considerations applied to $\Vec{v}\sub{e}\cdot\nabla\phi(\Vec{x},t)$, we can conclude that $-\Vec{v}\sub{e}\cdot\nabla f(\Vec{x},t)$ makes the vortex move from right to left in Fig. \ref{fig-vortex-moving} (c), and the third term in eq. (\ref{eq-single-vortex-dynamics}) describes the drag force parallel to $\Vec{v}\sub{e}$. These two forces enable us to investigate the effect of dissipation $\gamma(\Vec{x},t)$.

\begin{figure}[htb] \centering \begin{minipage}[t]{0.49\linewidth} \begin{center} \includegraphics[width=.99\linewidth]{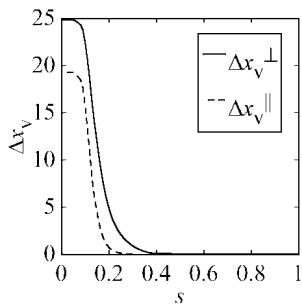} \end{center} \end{minipage}
\caption{\label{fig-s-dependence} Components of the vortex deviation $\Delta\Vec{x}\sub{v}$ for a range of $s$ values at $t=5$. In the calculation, $\gamma_0=1$ and $\gamma_0=0$ under the velocity field $\Vec{v}\sub{e}=0.5$.} \end{figure}

We numerically solved the GP equation (\ref{eq-vorticity-GP}) under the flow $\Vec{v}\sub{e}=0.5$ with a dissipation of $\gamma_0=1$ and no dissipation $\gamma_0=0$. The effect of dissipation can be determined by comparing the position of the vortex in two simulations. Here we let the dissipation have a scale-dependence as $\tilde{\gamma}(\Vec{k})=\gamma_0\theta(k-2s\pi/\xi)$ with the parameter $s$, and monitor how the vortex shift depends on $s$. The position of a vortex can be identified by the singularity around which the phase rotates by $2\pi$. Although the vortex just moves along $\Vec{v}\sub{e}$ for $\gamma_0=0$, the dissipation $\gamma_0=1$ changes its trajectory. The deviation $\Delta\Vec{x}\sub{v}(t)$ is divided into the part $\Delta x^{\perp}\sub{v}(t)$ perpendicular to $\Vec{v}\sub{e}$ and the part $\Delta x^{\parallel}\sub{v}$ parallel to $\Vec{v}\sub{e}$. Figure \ref{fig-s-dependence} shows how the deviation depends on $s$.
Both $\Delta x^{\perp}\sub{v}$ and $\Delta x^{\parallel}\sub{v}$ are negligible for $s\gtrsim 0.3$ and they rapidly increase for $s\lesssim 0.3$. From these considerations, we conclude that the dissipation with $s=1$ is adequate for the present study.

At finite temperatures, vortices in QT dissipate at various scales, which may be considered as the mutual friction of superfluid \He4. Also, the dependence of $\gamma(\Vec{x},t)$ on $s$ and $\gamma_0$ may correspond to the temperature-dependence of the mutual friction, but this possibility will not be determined here.

\section{Decaying Turbulence}\label{sec-decay}

In this section, we discuss a simulation of QT starting from a random phase profile \cite{Kobayashi}. With the dissipation term, the QT decayed gradually and fell into a quasi-steady turbulent state in which the vortices at large scales were nearly isotropic and uniformly. After showing that the decaying mechanism of QT is caused by the dissipation of compressible excitations, we obtain the energy spectrum of such a decaying QT and show that it is consistent with the Kolmogorov law.

\begin{figure}[htb] \centering \begin{minipage}{0.49\linewidth} \begin{center} \includegraphics[width=.99\linewidth]{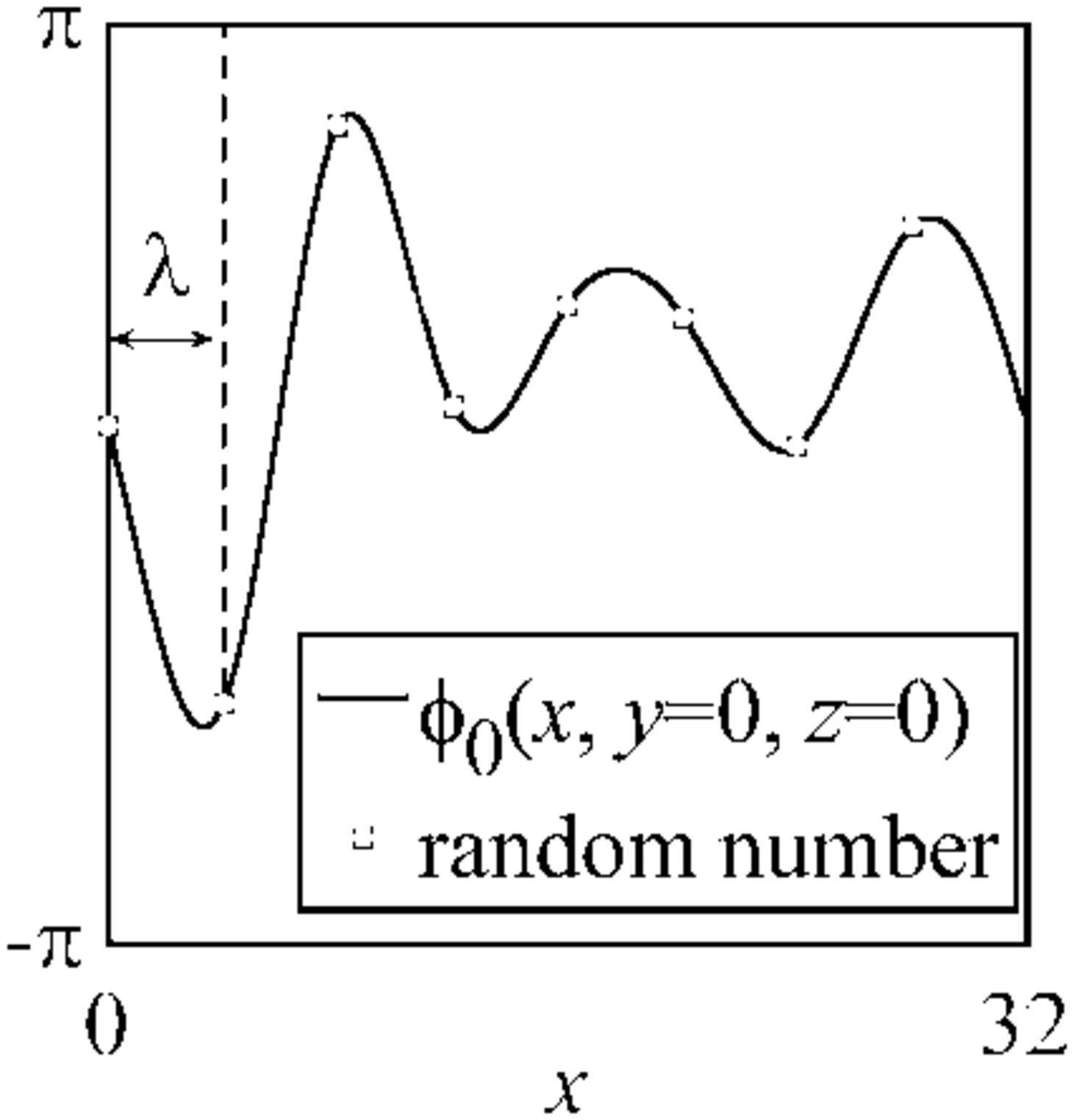}\\ (a) \end{center} \end{minipage} \begin{minipage}{0.49\linewidth} \begin{center} \includegraphics[width=.99\linewidth]{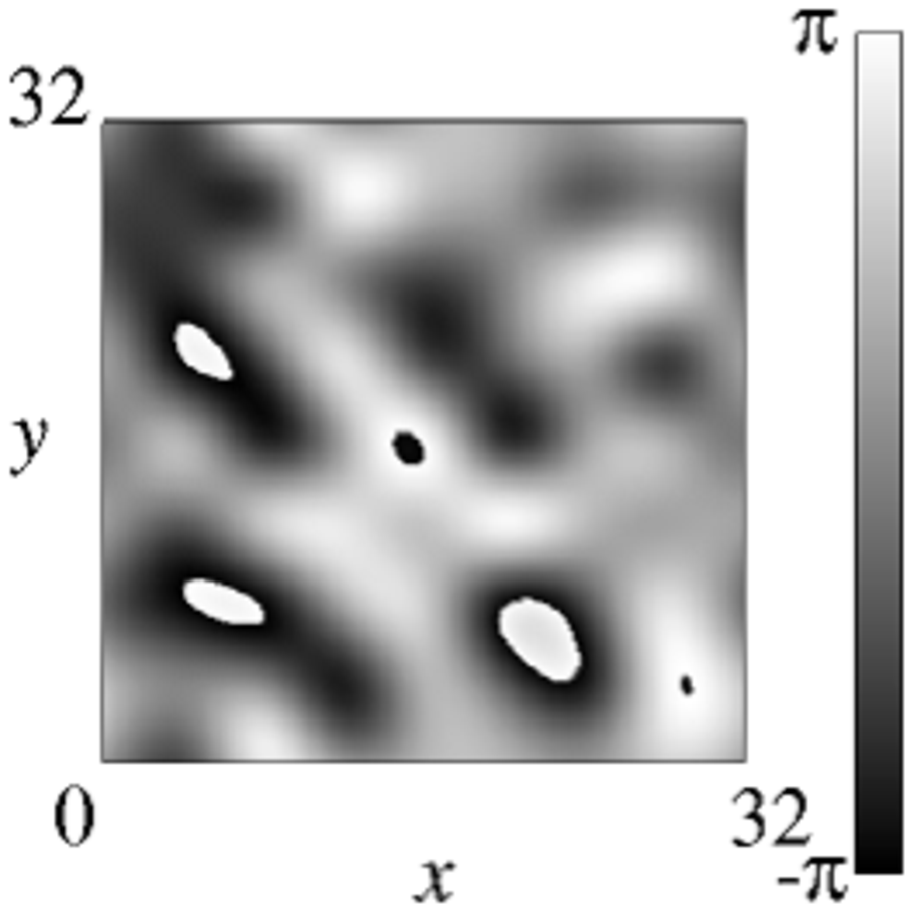}\\ (b) \end{center} \end{minipage}
\caption{\label{fig-random-phase} Procedure for making a random phase. (a) Example of $\phi_0(\Vec{x})$ with $\lambda=4$. (b) Example of an initial random phase in a cross-section $\phi_0(x,y,0)$.} \end{figure}

To obtain a turbulent state, we start from an initial configuration in which the density $f_0^2=1$ is uniform and the phase $\phi_0(\Vec{x})$ has a random spatial distribution. Here, the random phase $\phi_0(\Vec{x})$ is made by placing random numbers between $-\pi$ to $\pi$ at every distance $\lambda=4$ and connecting them smoothly (Fig. \ref{fig-random-phase}) by using a three-dimensional spline interpolation \cite{Press}.
The initial velocity $\Vec{v}(\Vec{x},t=0)=2\nabla\phi_0(\Vec{x})$ given by the initial random phase is random; as a result, the initial wave function is dynamically unstable and soon produces fully developed turbulence with many quantized vortex loops.

Before investigating the dynamics and statistics of QT, we discuss the numerical accuracy and the aliasing error. Because the aliasing error is maximized for $\gamma_0=0$, where the wave function with the high wave numbers is never dissipated, it would be enough to confirm that the aliasing error is very small for a simulation with $\gamma_0=0$. We can estimate the aliasing error by calculating the dynamics with a larger wave number space, which decreases the aliasing error, and then comparing the result with the dynamics of the original space. Such a large wave number space is realized by increasing the spatial resolution $N\sub{grid}=512^3$ ($\Delta x=0.0625$). For the two spaces, we compare the total energy $E(t)$, the interaction energy $E\sub{int}(t)$, the quantum energy $E\sub{q}(t)$, and the kinetic energy $E\sub{kin}(t)$ \cite{Nore}:
\begin{align} E(t)&=\frac{1}{N}\int\dd\Vec{x}\:\Phi^\ast(\Vec{x},t)\Big[-\nabla^2+\frac{g}{2}f(\Vec{x},t)^2\Big]\Phi(\Vec{x},t),\nonumber\\ E\sub{int}(t)&=\frac{g}{2N}\int\dd\Vec{x}\:f(\Vec{x},t)^4,\nonumber\\ E\sub{q}(t)&=\frac{1}{N}\int\dd\Vec{x}\:[\nabla f(\Vec{x},t)]^2,\nonumber\\ E\sub{kin}(t)&=\frac{1}{N}\int\dd\Vec{x}\:[f(\Vec{x},t)\nabla\phi(\Vec{x},t)]^2.\label{eq-energy-definition} \end{align}
In addition, we investigate the accuracy of the Runge-Kutta-Verner method for the time evolution by calculating the dynamics with higher time resolution $\Delta t=2\times 10^{-5}$. For the comparison, we estimate the relative errors $F_{12}(A)=|(\bracket{A}_1-\bracket{A}_2)/\bracket{A}_1|$ and $F_{13}(A)=|(\bracket{A}_1-\bracket{A}_3)/\bracket{A}_1|$ at $t=14$ for $A=E(t)$, $E\sub{int}(t)$, $E\sub{q}(t)$ and $E\sub{kin}(t)$. Here $\bracket{}_1$ means that the resolution was (i) $\Delta t=1\times 10^{-4}$ and $256^3$ grid points, $\bracket{}_2$ means a resolution of (ii) $\Delta t=2\times 10^{-5}$ and $256^3$ grid points, and $\bracket{}_3$ means (iii) $\Delta t=1\times 10^{-4}$ and $512^3$ grid points. The comparison in Table \ref{table-error} below shows that the relative errors are extremely small, which confirms the small aliasing error and the high accuracy of the time evolution, and allows us to use the original resolution of (i) above.
\begin{table}[htb] \begin{center} \caption{\label{table-error} Dependence of $F_{12}$ and $F_{13}$ on $E(t)$, $E\sub{int}(t)$, $E\sub{q}(t)$ and $E\sub{kin}(t)$.} \vspace*{0.4cm} \begin{tabular}{ccc} & $F_{12}(A)$ & $F_{13}(A)$ \\ \hline $A=E(t)$ & $2.5\times 10^{-12}$ & $6.5\times 10^{-10}$ \\ $A=E\sub{int}(t)$ & $3.9\times 10^{-12}$ & $9.0\times 10^{-10}$ \\ $A=E\sub{q}(t)$ & $2.6\times 10^{-12}$ & $7.1\times 10^{-10}$ \\ $A=E\sub{kin}(t)$ & $5.2\times 10^{-12}$ & $9.7\times 10^{-10}$ \\ \end{tabular} \end{center} \end{table}
The total energy is conserved to an accuracy of $10^{-10}$.

\begin{figure}[htb] \centering \begin{minipage}{0.49\linewidth} \begin{center} \includegraphics[width=.99\linewidth]{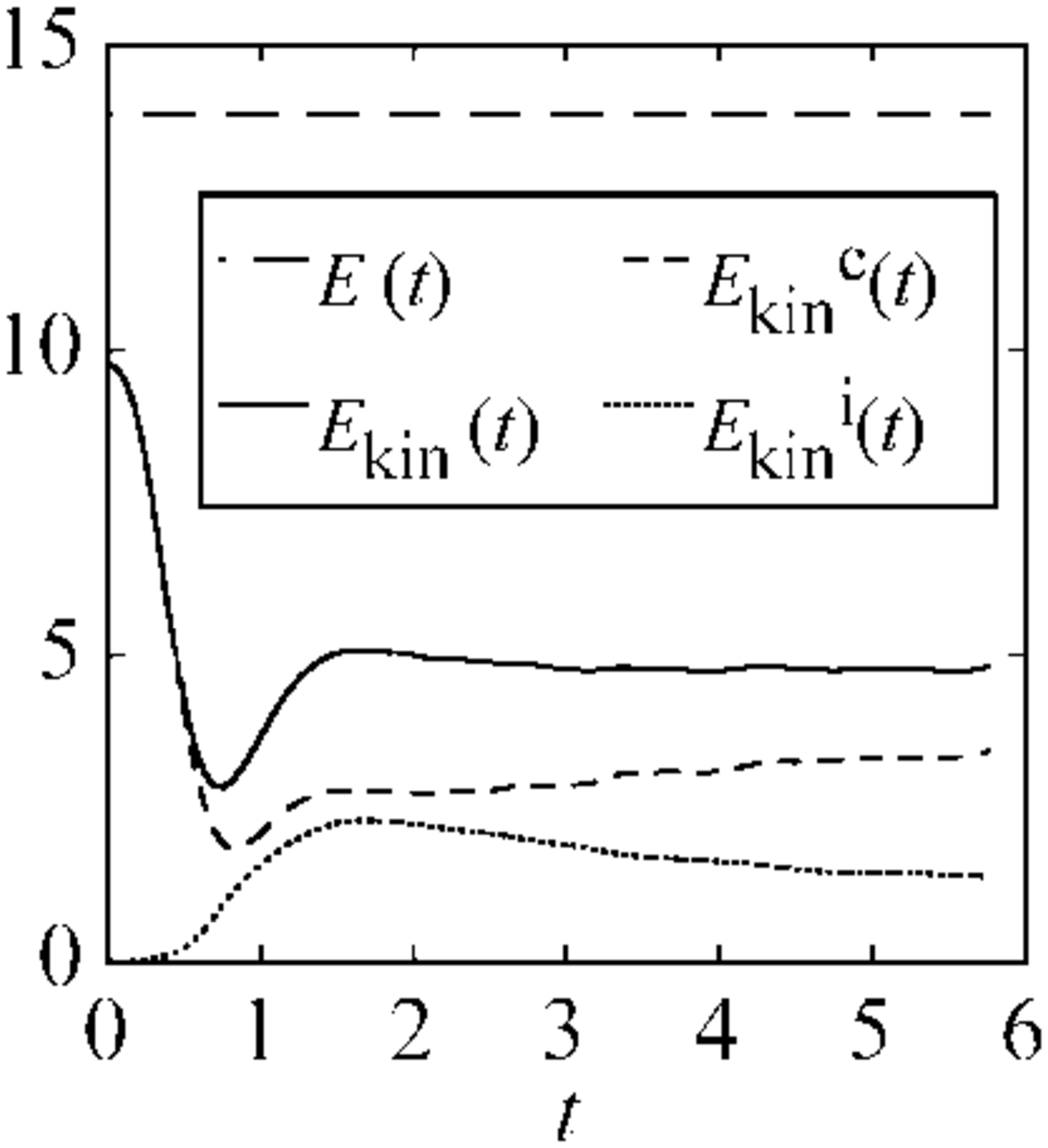}\\ (a) \end{center} \end{minipage} \begin{minipage}{0.49\linewidth} \begin{center} \includegraphics[width=.99\linewidth]{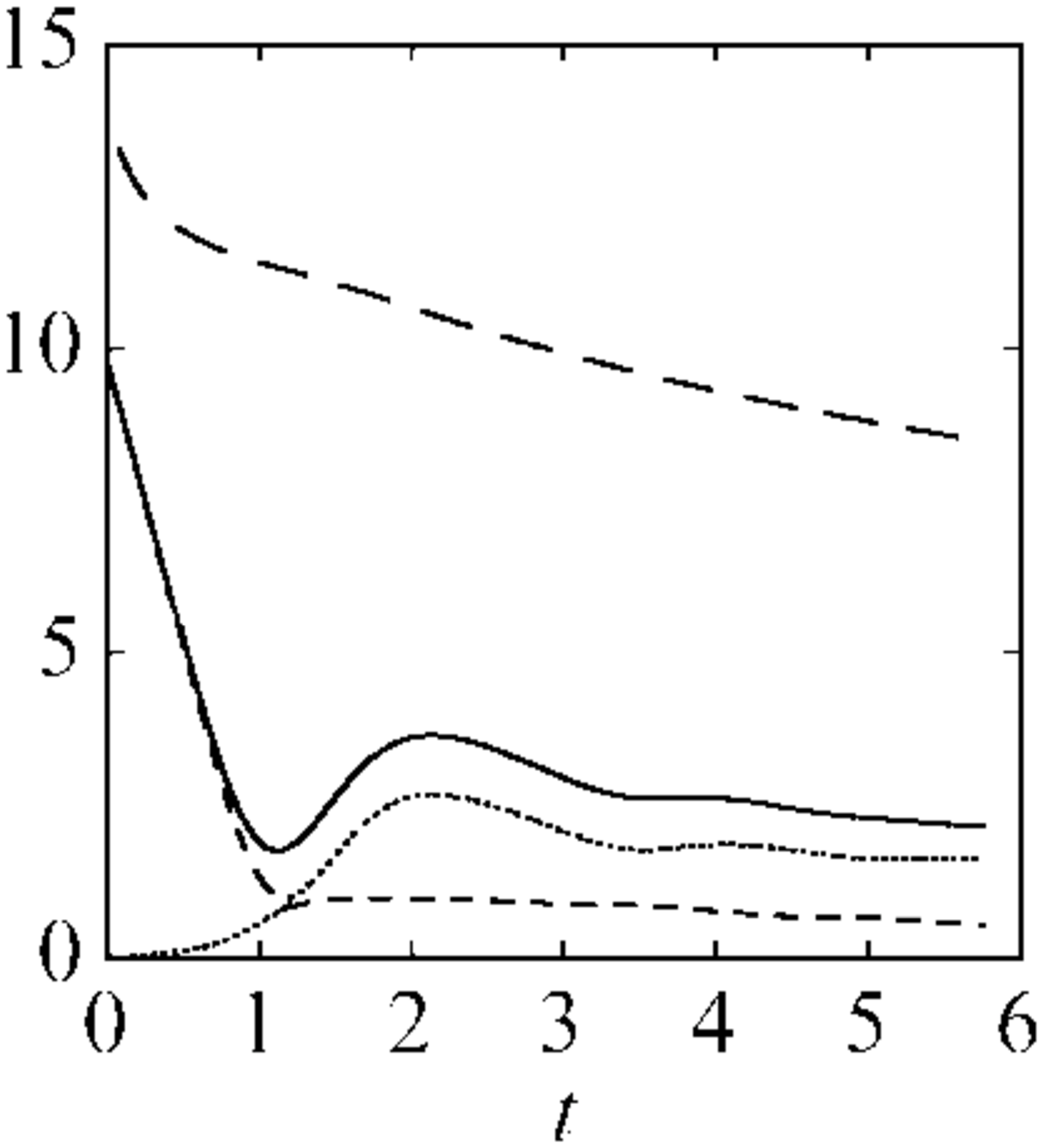}\\ (b) \end{center} \end{minipage}
\caption{\label{fig-gamma-energy} Time development of $E(t)$, $E\sub{kin}(t)$, $E\sub{kin}\up{c}(t)$, and $E\sub{kin}\up{i}(t)$. (a) Case with $\gamma_0=0$. (b) Case with $\gamma_0=1$ (Kobayashi and Tsubota, Phys. Rev. Lett. 94, 065302-3, 2005, reproduced with permission. Copyright (2005) by the American Physical Society).} \end{figure}

Next, we ran a simulation of QT with the dissipation $\gamma_0=1$ to remove compressible short-wavelength excitations. The effect of this dissipation is shown in Fig. \ref{fig-gamma-energy}. We divide up the kinetic energy $E\sub{kin}(t)$ into the compressible part $E\sub{kin}\up{c}(t)=\int\dd\Vec{x}\:[\{f(\Vec{x},t)\nabla\phi(\Vec{x},t)\}\up{c}]^2/N$, which is due to compressible excitations, and the incompressible part $E\sub{kin}\up{i}(t)=\int\dd\Vec{x}\:[\{f(\Vec{x},t)\nabla\phi(\Vec{x},t)\}\up{i}]^2/N$, which is due to vortices. Here $\{\cdots\}\up{c}$ denotes the compressible part $\nabla\times\{\cdots\}\up{c}=0$ and $\{\cdots\}\up{i}$ denotes the incompressible part $\nabla\cdot\{\cdots\}\up{i}=0$. Numerically, the compressible part $\Vec{A}\up{c}(\Vec{x})$ and the incompressible part $\Vec{A}\up{i}(\Vec{x})$ of an arbitrary vector field $\Vec{A}(\Vec{x})$ are given by
\begin{align} \Vec{A}\up{c}(\Vec{x})&=\sum_{\Vec{k}}\frac{\Vec{k}\cdot\tilde{\Vec{A}}(\Vec{k})}{k^2}\Vec{k}\exp(\ii\Vec{k}\cdot\Vec{x})\nonumber\\ \Vec{A}\up{i}(\Vec{x})&=\Vec{A}(\Vec{x})-\Vec{A}\up{c}(\Vec{x}),\label{eq-compressible} \end{align}
where $\tilde{\Vec{A}}(\Vec{k})$ is the Fourier component of $\Vec{A}(\Vec{x})$.
Figure \ref{fig-gamma-energy} shows the time development of $E(t)$, $E\sub{kin}(t)$, $E\sub{kin}\up{c}(t)$ and $E\sub{kin}\up{i}(t)$ for the cases of $\gamma_0=0$ (a) and $\gamma_0=1$ (b).
Without dissipation, the compressible kinetic energy $E\sub{kin}\up{c}(t)$ is increased in spite of the total energy conservation (Fig. \ref{fig-gamma-energy} (a)), a result that is consistent with the simulation by Nore {\it et al.} \cite{Nore} The dissipation suppresses $E\sub{kin}\up{c}(t)$ and thus causes $E\sub{kin}\up{i}(t)$ to be dominant in Fig. \ref{fig-gamma-energy} (b). This dissipation term acts as a viscosity term only at the small scales. At other scales, which are equivalent to the inertial range $\Delta k<k<2\pi/\xi$, the energy does not dissipate.

\begin{figure*}[htb] \centering \begin{minipage}{0.24\linewidth} \begin{center} \includegraphics[width=.99\linewidth]{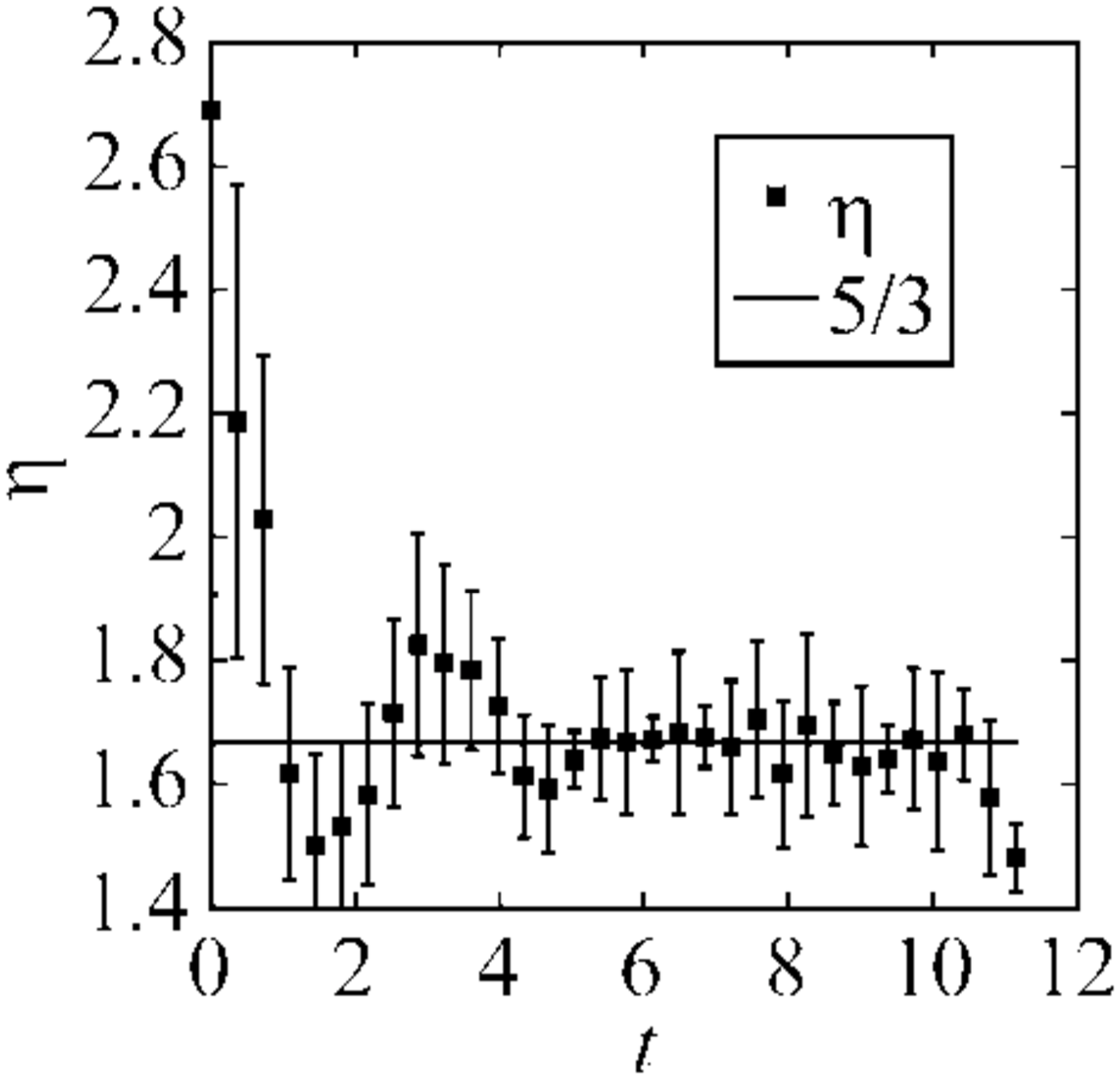}\\ (a) \end{center} \end{minipage} \begin{minipage}{0.24\linewidth} \begin{center} \includegraphics[width=.99\linewidth]{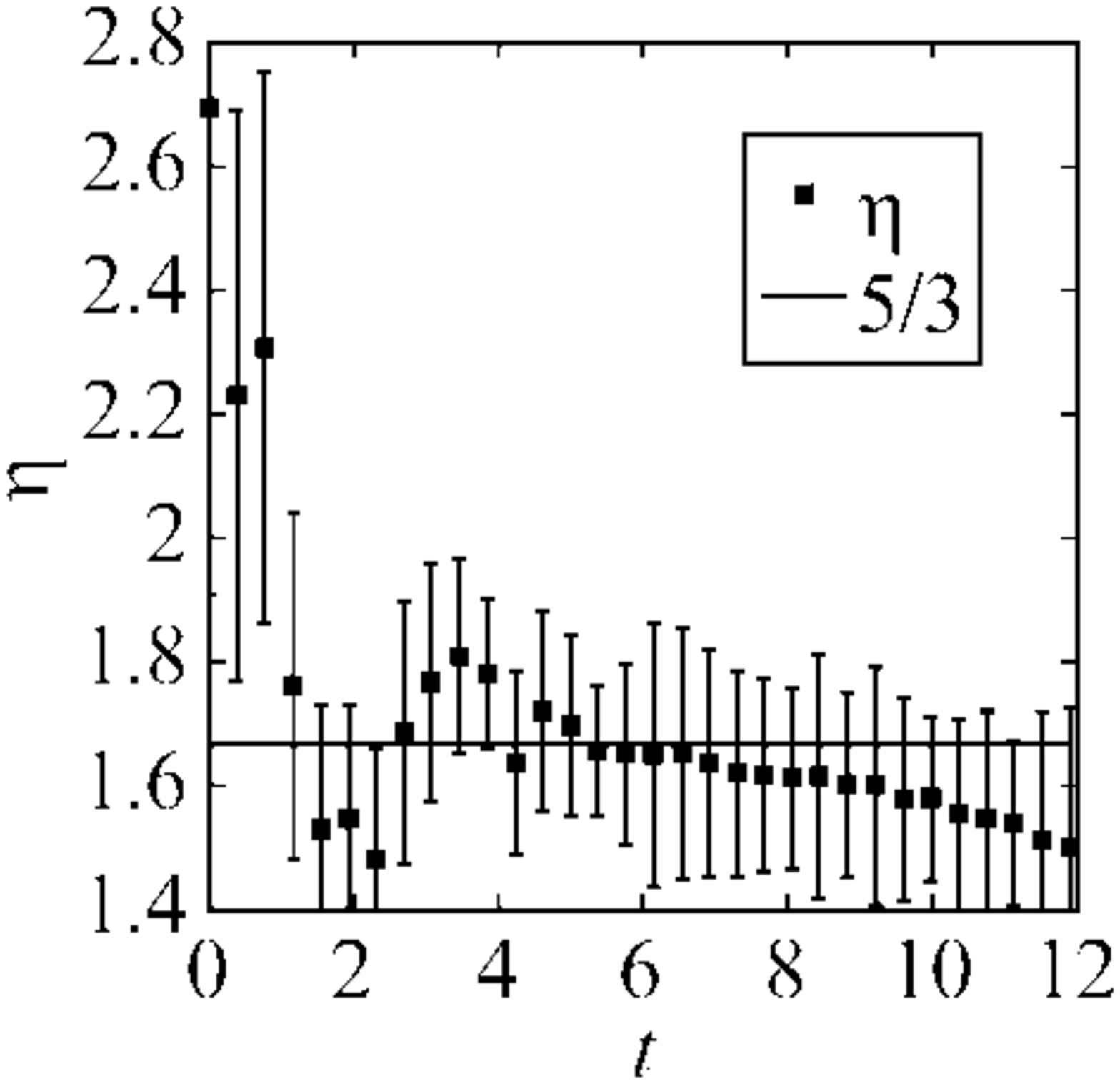}\\ (b) \end{center} \end{minipage} \begin{minipage}{0.24\linewidth} \begin{center} \includegraphics[width=.99\linewidth]{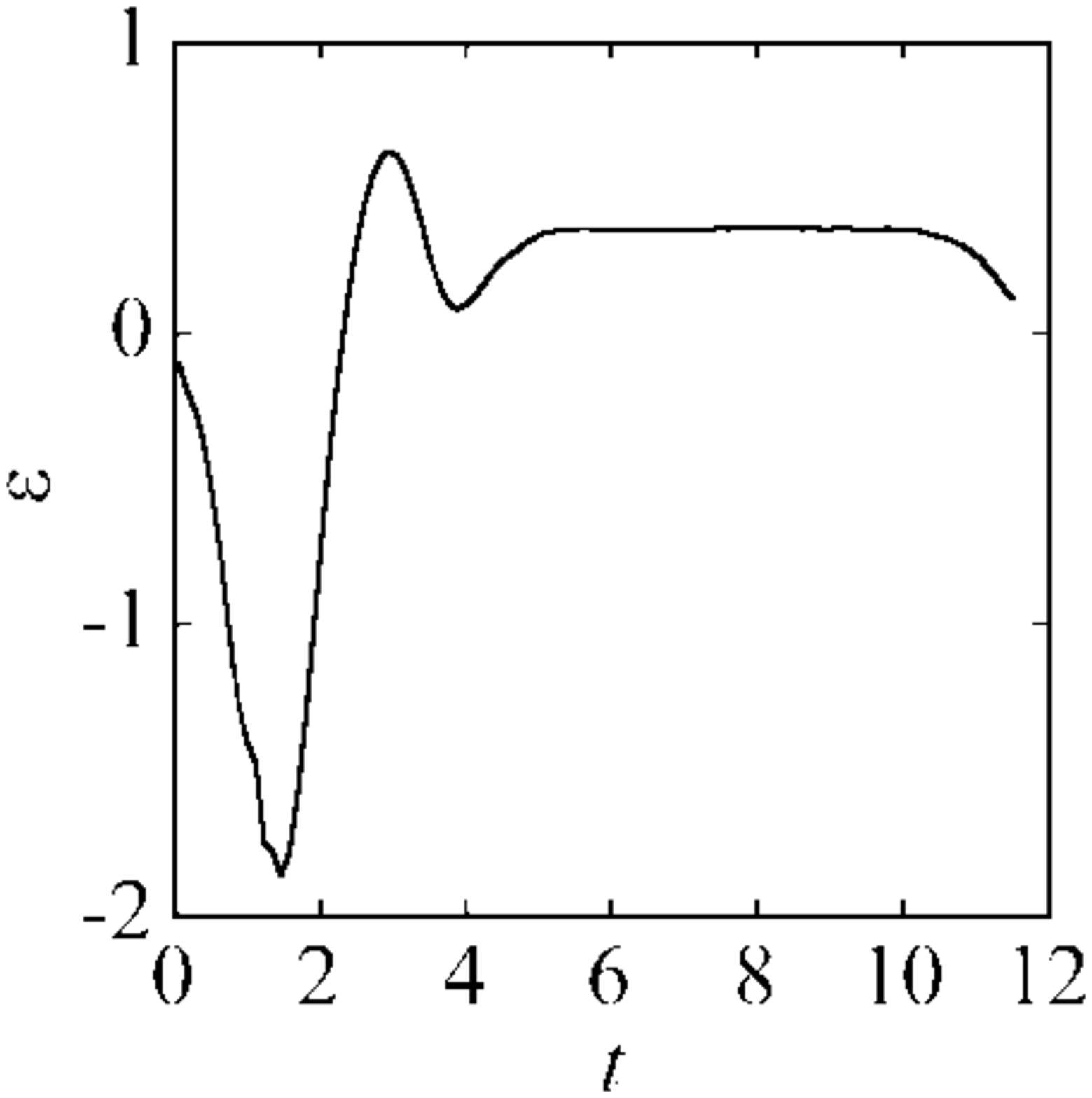}\\ (c) \end{center} \end{minipage} \begin{minipage}{0.24\linewidth} \begin{center} \includegraphics[width=.99\linewidth]{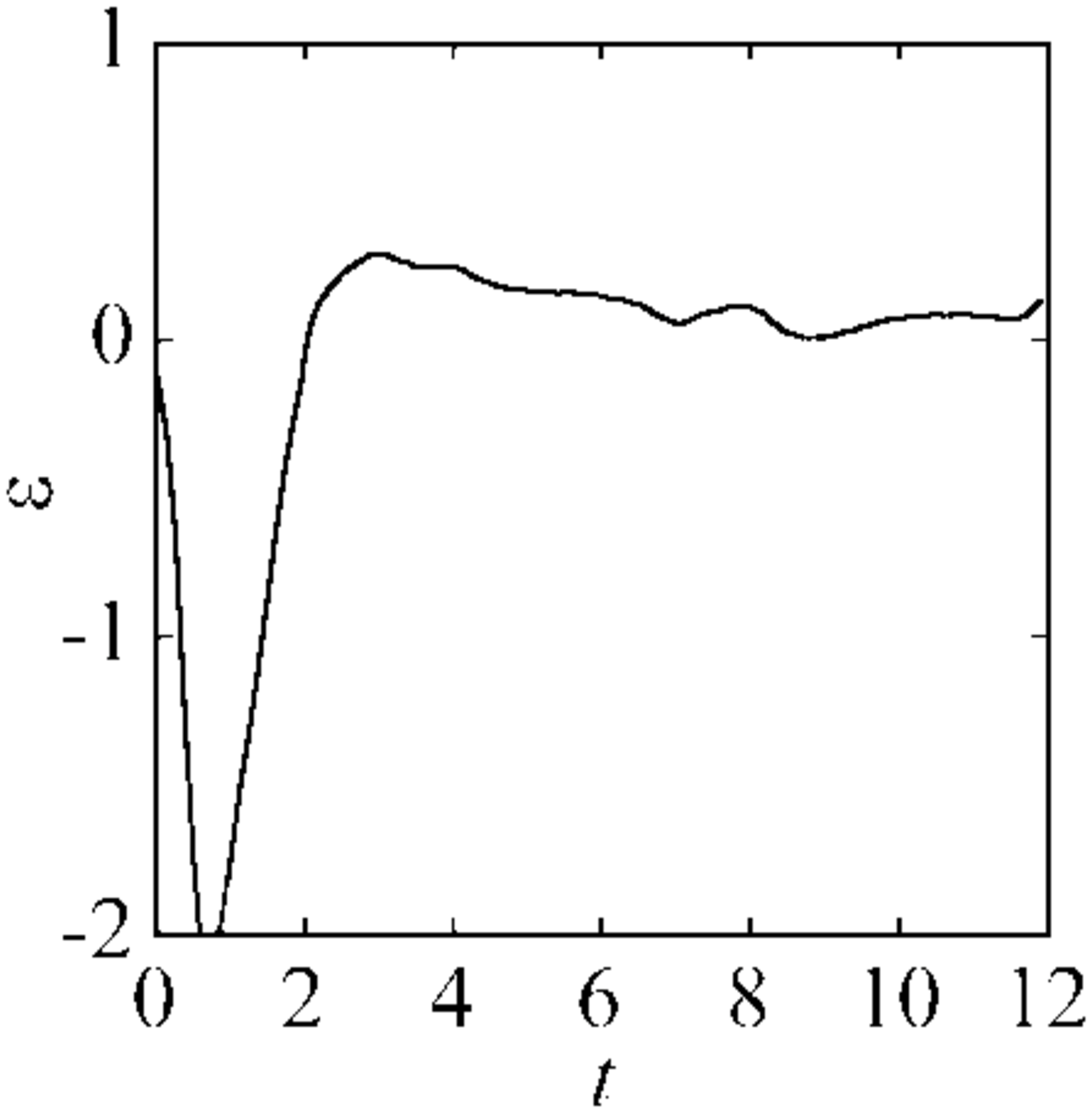}\\ (d) \end{center} \end{minipage}
\caption{\label{fig-exponent} Time dependence of the exponent of the spectrum $\eta$ and the dissipation rate $\varepsilon$. (a) $\eta$ for $\gamma_0=1$. (b) $\eta$ for $\gamma_0=0$. (c) $\varepsilon$ for $\gamma_0=1$. (d) $\varepsilon$ for $\gamma_0=0$. In (a) and (b), the error bars are the standard deviation of the data from the fit. The line $\eta=5/3$ is shown to compare the results with the Kolmogorov law (Kobayashi and Tsubota, Phys. Rev. Lett. 94, 065302-3, 2005, reproduced with permission. Copyright(2005) by the American Physical Society).} \end{figure*}

To show that the result is consistent with the Kolmogorov law, we calculate the spectrum of the incompressible kinetic energy $E\sub{kin}\up{i}(k,t)$ defined as $E\sub{kin}\up{i}(t)=\int\dd k\: E\sub{kin}\up{i}(k,t)$. Initially, the spectrum $E\sub{kin}\up{i}(k,t)$ significantly deviates from the Kolmogorov power-law; however, the spectrum approaches a power-law as the turbulence develops. We assume that the spectrum $E\sub{kin}\up{i}(k,t)$ is proportional to $k^{-\eta}$ in the inertial range $\Delta k<k<2\pi/\xi$ and then determine the exponent $\eta$ by making a straight fit to a log-log plot of the spectrum $E\sub{kin}\up{i}(k,t)$ only in the inertial range $\Delta k<k<2\pi/\xi$ at each time. The resulting time development of $\eta$ is shown in Fig. \ref{fig-exponent} (a).
This figure shows that the QT satisfies the Kolmogorov law $E\sub{kin}\up{i}(k,t)\propto k^{-5/3}$ for time $4\lesssim t\lesssim 10$, the times when the system may be fully developed turbulence. We also calculate the energy dissipation rate $\varepsilon=-\partial E\sub{kin}\up{i}(t)/\partial t$ and compare the results quantitatively with the Kolmogorov law. Figure \ref{fig-exponent} (c) shows that $\varepsilon$ is almost constant in the period $4\lesssim t\lesssim 10$, giving $\eta=5/3$, which means that the system can be considered to be a quasi-steady turbulent state with a constant Kolmogorov spectrum of $\varepsilon^{2/3}k^{-\eta}$.

Without the dissipation, that is, for $\gamma_0=0$, the QT did not satisfy the Kolmogorov law. This means that dissipating short-wavelength excitations are essential for producing the Kolmogorov law. Figure \ref{fig-exponent} (b) shows the time development of the exponent $\eta$ for $\gamma_0=0$. In this case, $\eta$ is consistent with the Kolmogorov law only in the short period $4\lesssim t\lesssim 7$. For $t\gtrsim 7$, the dynamics of quantized vortices are affected by active compressible excitations. Such an effect of compressible excitations is also clearly shown in Fig. \ref{fig-exponent} (d). Compared with Fig. \ref{fig-exponent} (c), $\varepsilon$ is unsteady and smaller than that of $\gamma_0=1$ and even becomes negative at times. When $\varepsilon$ is negative, the energy flows backward from compressible excitations to vortices, which may prevent the energy spectrum from satisfying the Kolmogorov law for $t\gtrsim 7$. This result is consistent with the simulation by Nore {\it et. al.} \cite{Nore} According to Fig. \ref{fig-exponent} (a), the agreement between the energy spectrum and the Kolmogorov law becomes weak at the later stage of $t\gtrsim 10$, which may be due to the following reasons. In the period $4\lesssim t\lesssim 10$, the energy dissipation rate is constant and the energy spectrum agrees with the Kolmogorov law, which may support the finding that the system is in a quasi-steady turbulent state, and the Richardson cascade process works well at large scales. The dissipation is caused mainly by the removal of short-wavelength excitations that had been emitted at vortex reconnections. However, the system at the late stage $t\gtrsim 10$, being no longer quasi-steady turbulent in large scales, has only small vortices after the Richardson cascade process. Such a state is not regarded as a turbulent state, and thus the energy spectrums inconsistent with the Kolmogorov law of $\eta=5/3$. Emissions of excitations through vortex reconnections hardly occurs, which greatly reduces the energy dissipation rate $\varepsilon$. We can avoid such a disagreement by adding energy injection to the simulation. The result is steady turbulence, which is discussed in the next section.

\begin{figure}[htb] \centering \begin{minipage}{0.49\linewidth} \begin{center} \includegraphics[width=.99\linewidth]{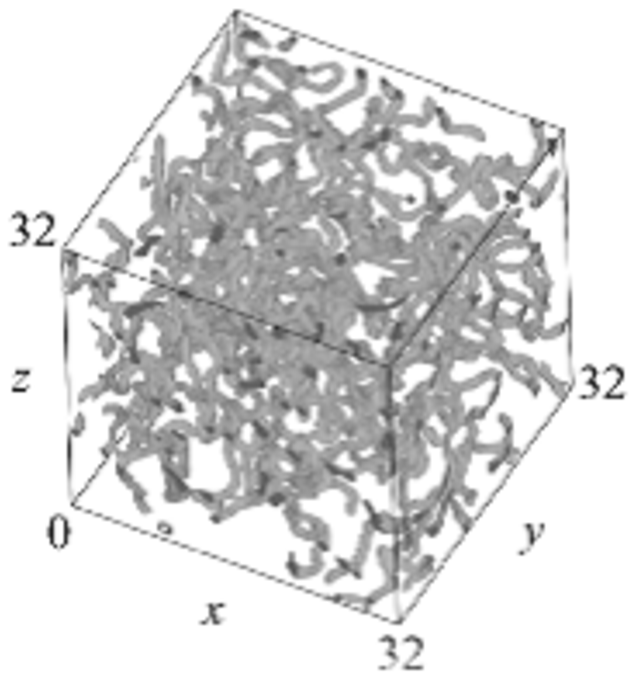}\\ (a) \end{center} \end{minipage} \begin{minipage}{0.49\linewidth} \begin{center} \includegraphics[width=.99\linewidth]{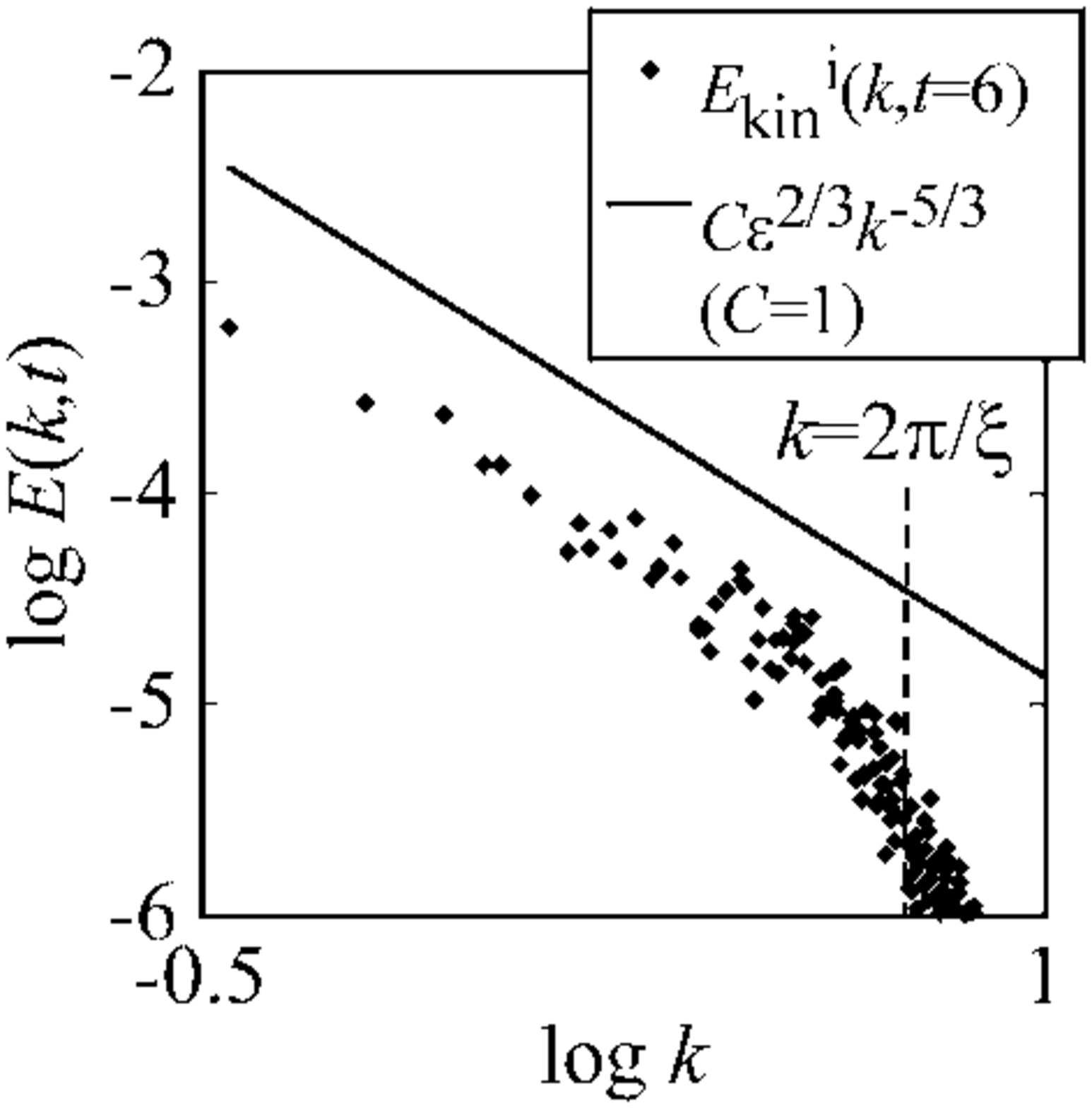}\\ (b) \end{center} \end{minipage}
\caption{\label{fig-decay-Kolmogorov} Simulated configuration of quantized vortices and the energy spectrum at $t=6$. (a) The configuration. (b) The energy spectrum, which was obtained by making an ensemble average of 20 initial states. The solid line refers to the Kolmogorov law in eq. (\ref{eq-Kolmogorov}).} \end{figure}

To confirm that the system is a fully developed turbulent state, we try to visualize the configuration of quantized vortices, the phase of which changes by $2\pi$ around the core. Because this is different from vortices in the vortex-filament model \cite{Schwarz, Araki}, identification of vortices in the GP model is not straightforward. Nevertheless, the identification can be done using the following two methods. The first is based on the fact that vortex cores have nearly infinite vorticity; in this method, we determine the points with a huge vorticity $|\Vec{\omega}(\Vec{x},t)|$. We used this method in our previous study \cite{Kobayashi}. The other method is to directly search for the points around which the phase rotates by $2\pi$. For simplicity, we consider a straight vortex line along the z-axis. At a point $(x,y,z)$ on the vortex core, the phases of the four surrounding points $(x+\Delta x,y,z)$, $(x,y+\Delta x,z)$, $(x-\Delta x,y,z)$, and $(x,y-\Delta x,z)$, all in the $x-y$ plane, rotate by $2\pi$. By extending this analysis to the $y-z$ and $z-x$ planes, we can find all points on a vortex core. These two methods of visualizing vortex cores are completely equivalent numerically, which means that quantized vortices are definite topological defects and thus different from eddies in classical fluid. An example of a fully developed turbulence with tangled vortices is shown in Fig \ref{fig-decay-Kolmogorov} (a). The system at this time ($t=6$) is in a quasi-steady turbulent state. Moreover, the energy spectrum $\varepsilon^{2/3}k^{-\eta}$ is in quantitative agreement with the Kolmogorov law (Fig. \ref{fig-decay-Kolmogorov} (b)). We thus conclude that the QT in quasi-steady state has the same statistics as the steady CT that follows the Kolmogorov law.
The inertial range $\Delta k<k<2\pi/\xi$ is now $0.2\lesssim k \lesssim 6.3$, which is wider than that in the simulations by Araki {\it et al.} \cite{Araki} and Nore {\it et al.} \cite{Nore}. The Kolmogorov constant is estimated to be $C=0.32\pm 0.1$, and smaller than that in CT which is estimated to be $1.4\lesssim C\lesssim 1.8$ \cite{Kida}. Araki {\it et al}. got the Kolmogorov constant $C=0.7$ in their numerical simulation of the vortex-filament model and this is also smaller than that in CT \cite{Araki}. This small Kolmogorov constant may be, therefore, characteristic of QT.

\section{Steady Turbulence}\label{sec-steady}

\begin{figure}[htb] \centering \begin{minipage}[t]{0.95\linewidth} \begin{center} \includegraphics[width=.99\linewidth]{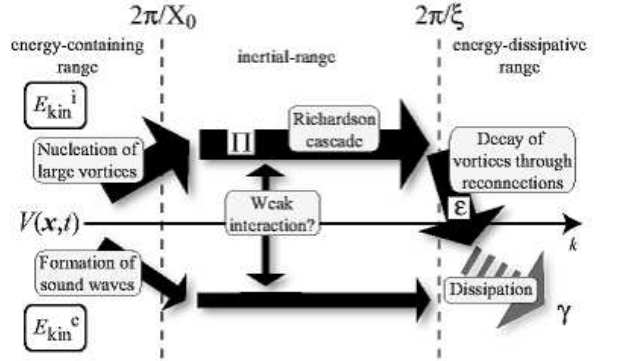} \end{center} \end{minipage} 
\caption{\label{fig-energy-current} The flow of the incompressible kinetic energy $E\sub{kin}\up{i}(t)$ (upper half of diagram) and compressible kinetic energy $E\sub{kin}\up{c}(t)$ (lower half) in wave number space.} \end{figure}

By introducing energy injection to the QT through a moving random potential, we model steady QT. The advantage of steady QT over decaying QT are the following. First, steady QT gives a clearer correspondence with the Kolmogorov law; this is because the original statistics in the inertial range has been developed for steady turbulence. Second, it enables us to confirm the presence of the energy-containing range, the inertial range and the energy-dissipative range of QT, which also exist in CT. Third, in all ranges, we can obtain the time-independent energy flux in the wave number space. Consequently, it is possible to reveal the cascade process of QT, which occur only by quantized vortices and short-wavelength excitations. This is later shown in Fig. \ref{fig-energy-current}.

For energy injection at large scales, we introduce the moving random potential $V(\Vec{x},t)$,
\begin{align} &[\ii-\gamma(\Vec{x},t)]\frac{\partial}{\partial t}\Phi(\Vec{x},t)\nonumber\\ &\quad =[-\nabla^2-\mu(t)+g|\Phi(\Vec{x},t)|^2+V(\Vec{x},t)]\Phi(\Vec{x},t).\label{eq-initial-GP-injection} \end{align}
We can numerically calculate the time development of the GP equation (\ref{eq-initial-GP-injection}) by the same procedure that is discussed in Sec. \ref{sec-procedure} except $\tilde{h}(\Vec{k},t)$ in eq. (\ref{eq-pseudo-Fourier-GP}) is replaced by the Fourier transformation of $h(\Vec{x},t)=[g|\Phi(\Vec{x},t)|^2+V(\Vec{x},t)]\Phi(\Vec{x},t)$. To make the moving random potential $V(\Vec{x},t)$, we used a similar method as that used for making the random phase $\phi_0(\Vec{x})$ in Sec. \ref{sec-decay} as the initial state of the wave function $\Phi(\Vec{x},t=0)=f_0\exp[\ii\phi_0(\Vec{x})]$. We place random numbers between $0$ and $V_0$ in space-time (\Vec{x},t) at every distance $X_0$ for space and $T_0$ for time, and connect them smoothly by using the four-dimensional spline interpolation \cite{Press}. We then obtain the moving random potential that has the following Gaussian two-point correlation
\begin{align} &\bracket{V(\Vec{x},t)V(\Vec{x}^{\prime},t^{\prime})}\nonumber\\ &\quad =V_0^2\exp\Big[-\frac{(\Vec{x}-\Vec{x}^{\prime})^2}{2X_0^2}-\frac{(t-t^{\prime})^2}{2T_0^2}\Big].\label{eq-random-correlation} \end{align}
This moving random potential has the characteristic spatial scale $X_0$ and thus quantized vortices of radius $X_0$ are nucleated when $V_0$ is strong enough. We define the wave number separating the energy-containing range and the inertial range as $2\pi/X_0$. The wave number $2\pi/\xi$ between the inertial range and the energy-dissipative range is defined by the dissipation term $\gamma(\Vec{x},t)$ the same as that in decaying turbulence. Therefore, our steady QT has the energy-containing range $k<2\pi /X_0$, the inertial range $2\pi/X_0<k<2\pi/\xi$ and the energy-dissipative range $2\pi/\xi <k$.

\begin{figure}[htb] \centering \begin{minipage}{0.49\linewidth} \begin{center} \includegraphics[width=.99\linewidth]{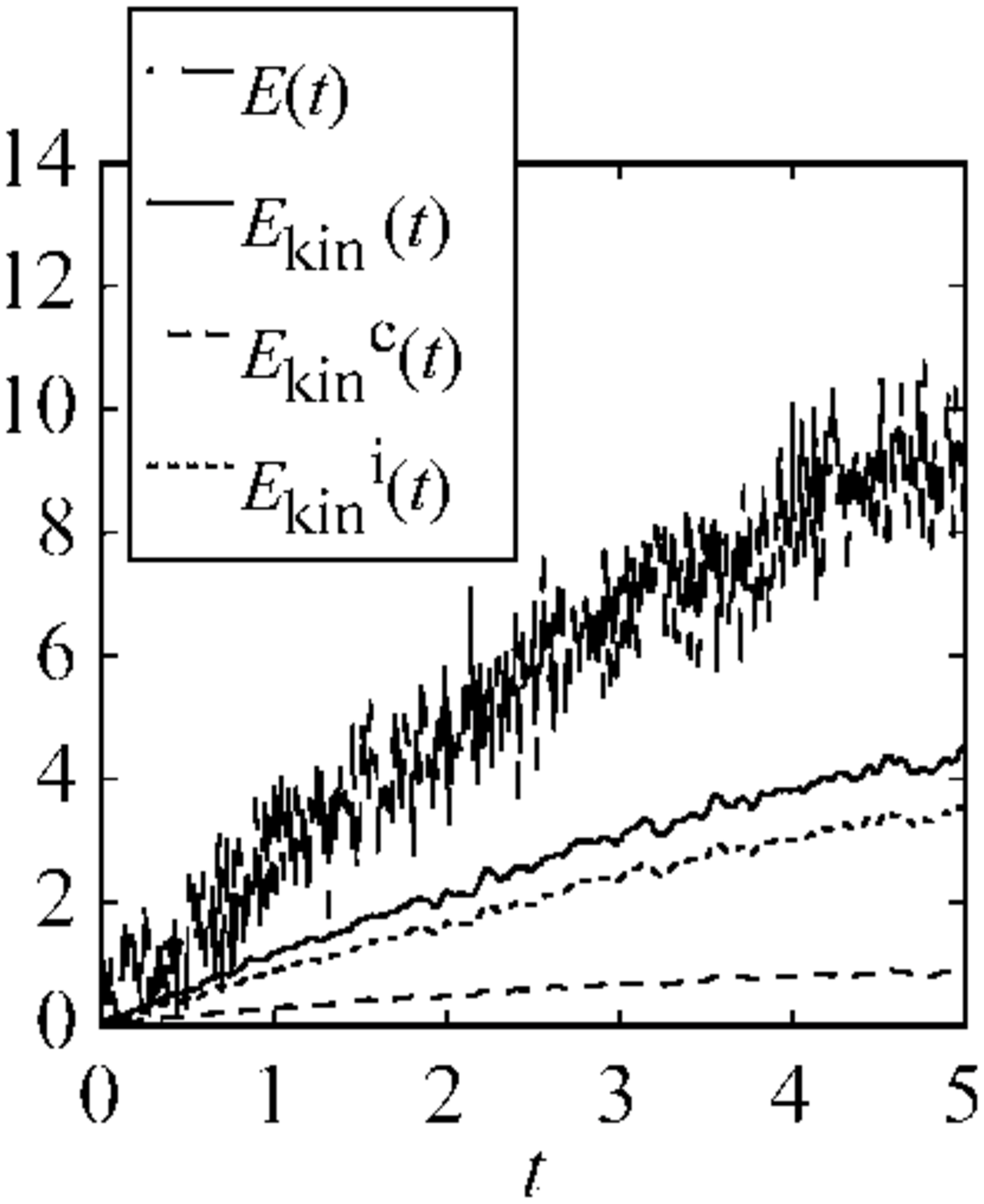}\\ (a) \end{center} \end{minipage} \begin{minipage}{0.49\linewidth} \begin{center} \includegraphics[width=.99\linewidth]{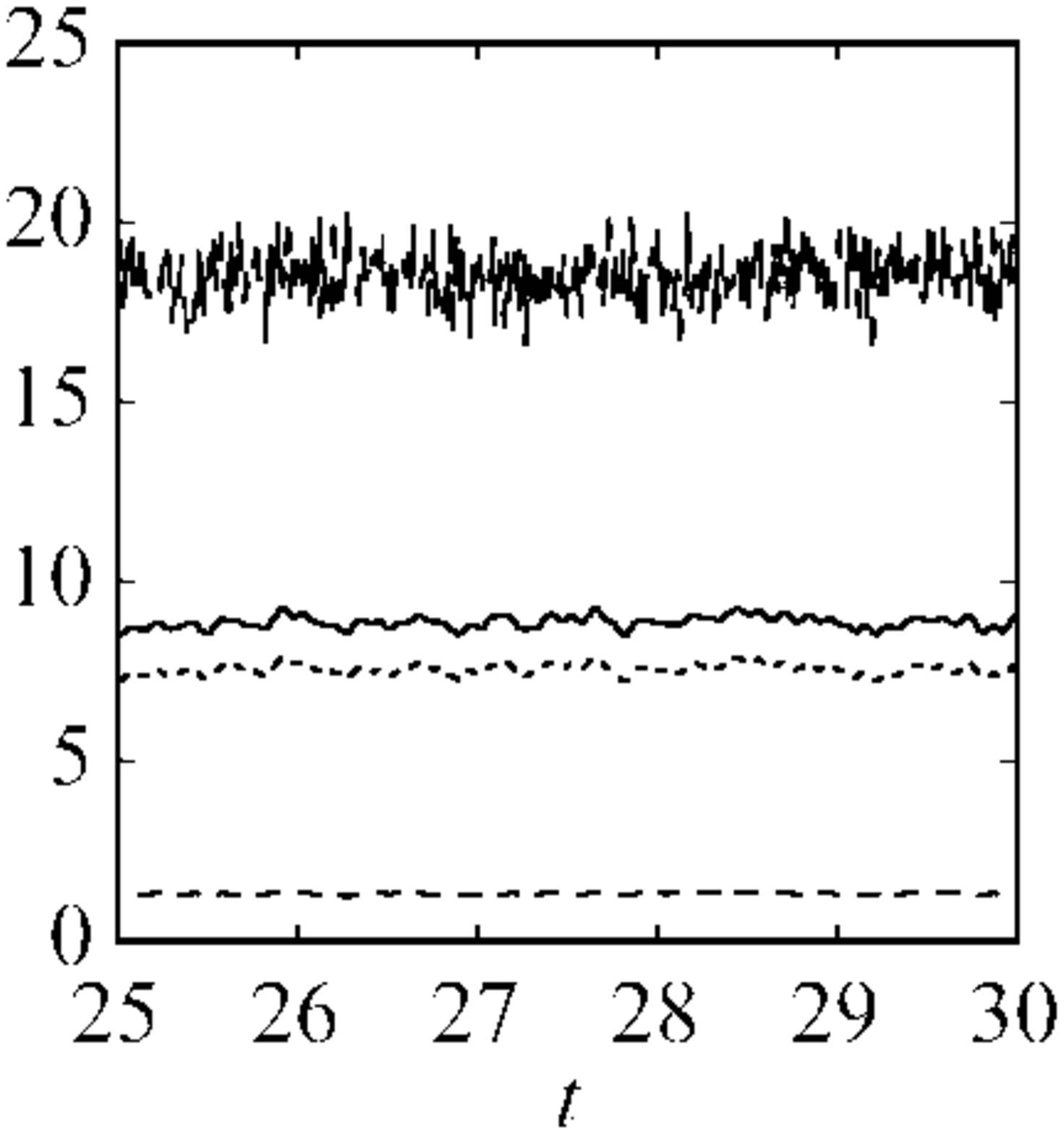}\\ (b) \end{center} \end{minipage}
\caption{\label{fig-steady-energy} Time development of $E(t)$, $E\sub{kin}(t)$, $E\sub{kin}\up{c}(t)$ and $E\sub{kin}\up{i}(t)$ at (a) the initial stage $0\le t\le 5$ and (b) a later stage $25\le t\le 30$ (b).} \end{figure}

We now show the result of our numerical simulation of steady turbulence for $V_0=50$, $X_0=4$, and $T_0=6.4\times10^{-2}$ which are suitable numerical parameters to create fully developed turbulence. We start from the uniform wave function $\Phi(\Vec{x},t=0)=f_0=1$. Figure \ref{fig-steady-energy} shows the time development of the total energy $E(t)$, the kinetic energy $E\sub{kin}(t)$, the compressible kinetic energy $E\sub{kin}\up{c}(t)$, and the incompressible kinetic energy $E\sub{kin}\up{i}(t)$.
Both figures show that the incompressible kinetic energy $E\sub{kin}\up{i}(t)$ is always dominant in the total kinetic energy $E\sub{kin}(t)$; the moving random potential contributes to the nucleation of vortices rather than that of sound waves. As shown in Fig. \ref{fig-steady-energy} (b), the four energies are almost constant for $t\gtrsim 25$; hence, we can obtain a steady QT after $t\simeq 25$. The vortex configuration of steady QT looks similar to that of decaying QT in Fig. \ref{fig-decay-Kolmogorov} (a).

This steady turbulence is sustained by the competition between energy injection and energy dissipation, which also results in a steady distribution of vortices. During steady QT, the energy flow in the wave number space looks like that in Fig. \ref{fig-energy-current}. The upper half of the diagram shows the kinetic energy $E\sub{kin}\up{i}(t)$ of vortices and the lower half shows the kinetic energy $E\sub{kin}\up{c}(t)$ of compressible excitations.
In the energy-containing range $k<2\pi /X_0$, the system receives incompressible kinetic energy from the moving random potential with the nucleation of large quantized vortex loops. Throughout the Richardson cascade process of these quantized vortices, the energy flows from small to large $\Vec{k}$ in the inertial range $2\pi/X_0<k<2\pi/\xi$. In the energy-dissipative range $2\pi/\xi <k$, the incompressible kinetic energy transforms to compressible kinetic energy through reconnections of vortices and the disappearance of small vortex loops. The moving random potential also creates long-wavelength compressible sound waves that are another source of compressible kinetic energy and also produce an interaction with vortices. However, the sound waves have a very weak effect because $E\sub{kin}\up{i}(t)$ is much larger than $E\sub{kin}\up{c}(t)$ as shown in Fig. \ref{fig-steady-energy}; moreover, sound waves of long wavelength have small scattering cross-section with vortex cores of the order of $\xi$.

\begin{figure}[htb] \centering \begin{minipage}{0.49\linewidth} \begin{center} \includegraphics[width=.99\linewidth]{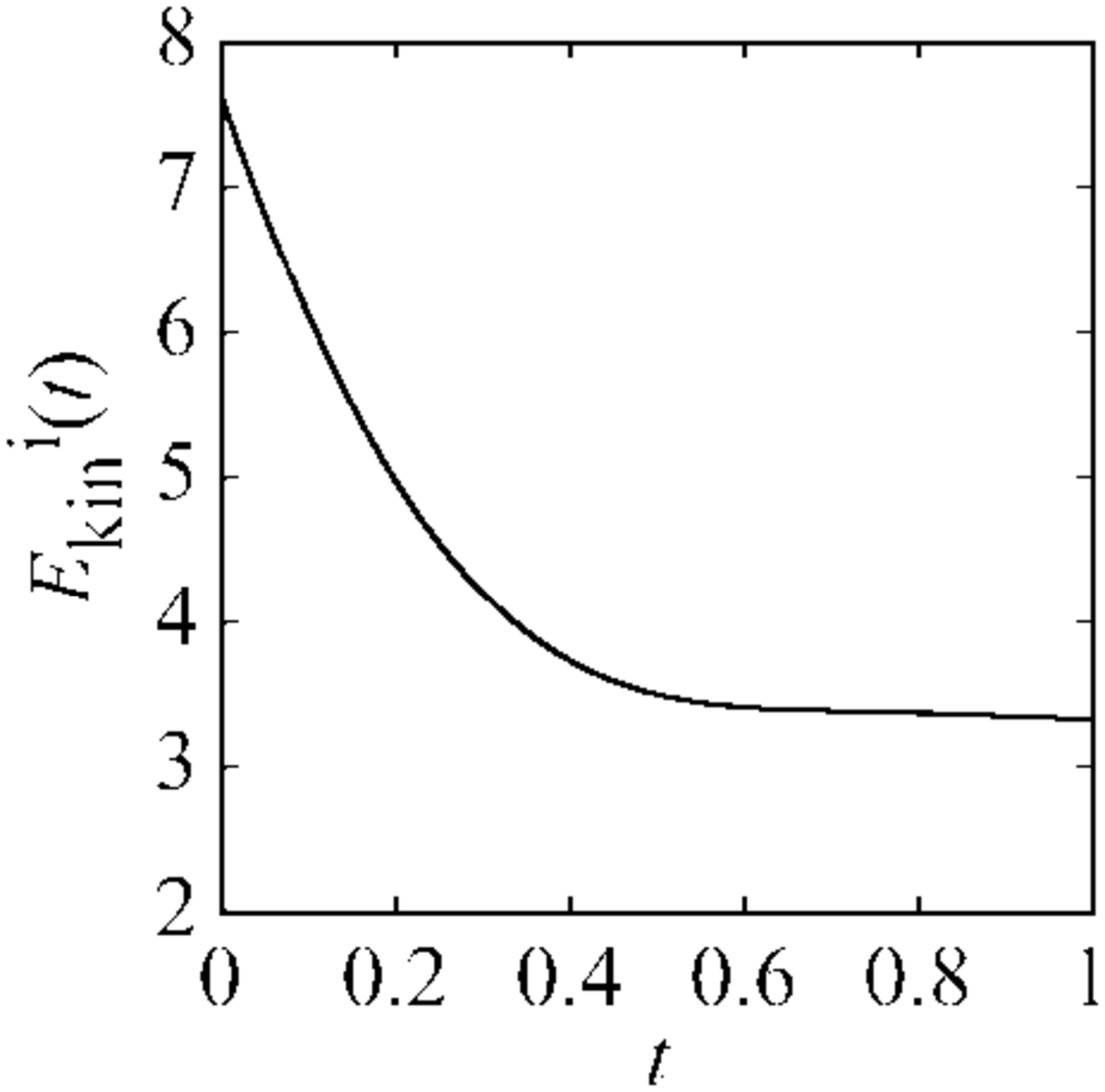}\\ (a) \end{center} \end{minipage} \begin{minipage}{0.49\linewidth} \begin{center} \includegraphics[width=.99\linewidth]{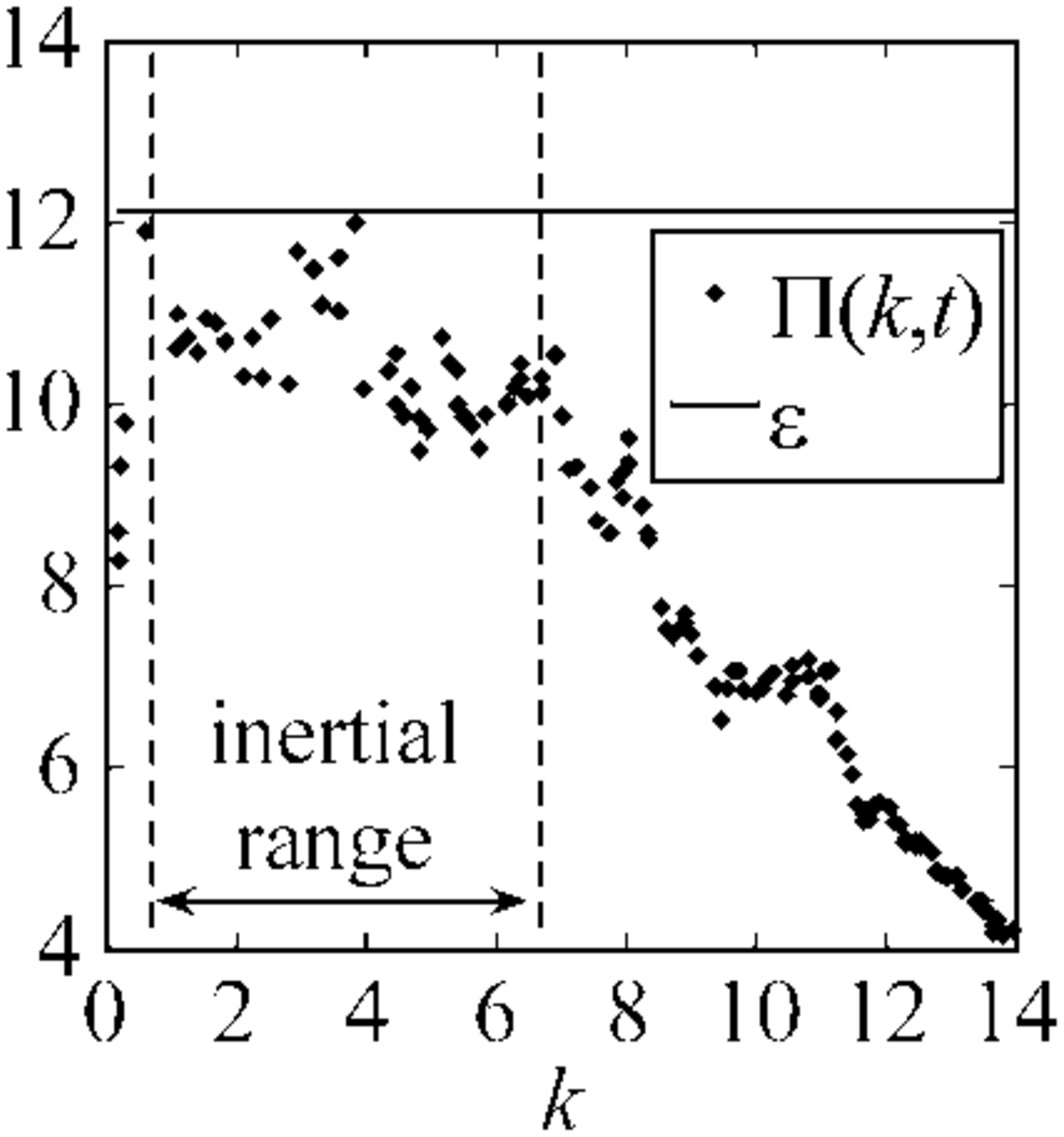}\\ (b) \end{center} \end{minipage}
\caption{\label{fig-energy-flux} (a) Time development of the incompressible kinetic energy $E\sub{kin}\up{i}(t)$ after switching off the moving random potential in a steady QT. (b) Dependence of the energy flux $\Pi(k,t)$ on the wave number $k$. The energy flux was calculated from an ensemble average of 50 randomly selected states at $t>25$. The solid line is the energy dissipation rate $\varepsilon$ calculated from Figure (a).} \end{figure}

To further test the scenario in Fig. \ref{fig-energy-current}, we investigate the energy dissipation rate $\varepsilon$ of $E\sub{kin}\up{i}(t)$ and the flux of energy $\Pi(k,t)$ through the Richardson cascade in the inertial range. The energy dissipation rate $\varepsilon$ of $E\sub{kin}\up{i}(t)$ in a steady QT can be equated to $\varepsilon=-\partial E\sub{kin}\up{i}(t)/\partial t$ after switching off the moving random potential. This is because the incompressible kinetic energy $E\sub{kin}\up{i}$ decays to the energy of compressible short-wavelength excitations. Figure \ref{fig-energy-flux} (a) shows one example of the time development of $E\sub{kin}\up{i}(t)$ after switching off the moving random potential. At the beginning, $E\sub{kin}\up{i}(t)$ decreases almost linearly at a rate from which we obtained $\varepsilon\simeq 12.5\pm 2.3$.
On the other hand, the energy flux $\Pi(k,t)$ of the incompressible kinetic energy in QT can be calculated by considering the scale-by-scale energy budget equation as well as CT \cite{Frisch}. To do this, we start from eq. (\ref{eq-Madelung-GP}) with the moving random potential $V(\Vec{x},t)$:
\begin{align} &f(\Vec{x},t)\frac{\partial}{\partial t}\phi(\Vec{x},t)\nonumber\\ &\quad =\gamma(\Vec{x},t)[2\nabla f(\Vec{x},t)\cdot\nabla\phi(\Vec{x},t)+f(\Vec{x},t)\nabla^2\phi(\Vec{x},t)]\nonumber\\ &\qquad +\nabla^2f(\Vec{x},t)-f(\Vec{x},t)[\nabla\phi(\Vec{x},t)]^2\nonumber\\ &\qquad -[gf(\Vec{x},t)^2-\mu(t)+V(\Vec{x},t)]f(\Vec{x},t),\label{eq-Pi-middle-1} \end{align}
and again neglect the terms $\nabla f(\Vec{x},t)$ and $\partial f(\Vec{x},t)/\partial t$ because they are not effective at scales exceeding $\xi$. Introducing a new variable $\Vec{p}(\Vec{x},t)=f(\Vec{x},t)\nabla\phi(\Vec{x},t)$, this equation reduces to
\begin{align} &\frac{\partial}{\partial t}\Vec{p}(\Vec{x},t)+\Vec{p}(\Vec{x},t)\cdot\nabla\Vec{v}(\Vec{x},t)\nonumber\\ &\quad =\gamma(\Vec{x},t)\nabla[\nabla\cdot\Vec{p}(\Vec{x},t)]+\Vec{F}(\Vec{x},t),\label{eq-Pi-middle-2} \end{align}
where $\Vec{F}(\Vec{x},t)=-\nabla[gf(\Vec{x},t)^2-\mu(t)+V(\Vec{x},t)]f(\Vec{x},t)$. Next, for any function $A(\Vec{x})$, we introduce the low-pass and high-pass filtered functions $A_k^<(\Vec{x})=\sum_{q\le k}\tilde{A}(\Vec{q})\exp[\ii\Vec{q}\cdot\Vec{x}]$ and $A_k^>(\Vec{x})=\sum_{q>k}\tilde{A}(\Vec{q})\exp[\ii\Vec{q}\cdot\Vec{x}]$ respectively, and the operator of the low-pass filter $L_k:A(\Vec{x})\to A_k^<(\Vec{x})$. Applying $L_k$ and taking the scalar product with $\Vec{p}^<_k(\Vec{x},t)$ to eq. (\ref{eq-Pi-middle-2}), we obtain
\begin{align} &\frac{1}{2}\frac{\partial}{\partial t}[\Vec{p}^<_k(\Vec{x},t)]^2+\Vec{p}^<_k(\Vec{x},t)\cdot L_k[\Vec{p}(\Vec{x},t)\cdot\nabla\Vec{v}(\Vec{x},t)]\nonumber\\ &\quad =\gamma(\Vec{x},t)\Vec{p}^<_k(\Vec{x},t)\cdot\nabla[\nabla\cdot\Vec{p}^<_k(\Vec{x},t)]\nonumber\\ &\qquad +\Vec{p}^<_k(\Vec{x},t)\cdot\Vec{F}^<_k(\Vec{x},t).\label{eq-Pi-middle-3} \end{align}
Extracting the incompressible part of eq. (\ref{eq-Pi-middle-3}) and integrating it over space yields the scale-by-scale incompressible kinetic energy budget equation
\begin{equation} \frac{\partial}{\partial t}\mathscr{E}(k,t)+\Pi(k,t)=-\Omega(k,t)+\mathscr{F}(k,t).\label{eq-energy-budget} \end{equation}
Here we introduce the cumulative incompressible kinetic energy $\mathscr{E}(k,t)$ between $0$ and $k$
\begin{equation} \mathscr{E}(k,t)=\frac{1}{N}\int\dd\Vec{x}\:[\{\Vec{p}^<_k(\Vec{x},t)\}\up{i}]^2,\label{eq-cumulative-energy} \end{equation}
the cumulative energy dissipation $\Omega(k,t)$
\begin{equation} \Omega(k,t)=-\frac{2}{N}\int\dd\Vec{x}\:\gamma(\Vec{x},t)\{\Vec{p}^<_k(\Vec{x},t)\cdot\nabla[\nabla\cdot\Vec{p}^<_k(\Vec{x},t)]\}\up{i},\label{eq-cumulative-dissipation} \end{equation}
the cumulative energy injection $\mathscr{F}(k,t)$
\begin{equation} \mathscr{F}(k,t)=\frac{2}{N}\int\dd\Vec{x}\:\{\Vec{p}^<_k(\Vec{x},t)\cdot F^<_k(\Vec{x},t)\}\up{i},\label{eq-cumulative-injection} \end{equation}
and the energy flux through $k$ of the incompressible kinetic energy
\begin{align} &\Pi(k,t)\nonumber\\ &\quad =\frac{2}{N}\int\dd\Vec{x}[\{\Vec{p}^<_k(\Vec{x},t)\cdot L_k[\Vec{p}(\Vec{x},t)\cdot\nabla\Vec{v}(\Vec{x},t)]\}\up{i}].\label{eq-Pi} \end{align}
Equation (\ref{eq-energy-budget}) can be interpreted as follows: at a given scale $k$, the rate of change of the incompressible kinetic energy is equal to the energy injected by the force $\mathscr{F}(k,t)$ minus the energy dissipation $\Omega(k,t)$ minus the energy flux $\Pi(k,t)$ to smaller scales. The calculated energy flux $\Pi(k,t)$ in Fig. \ref{fig-energy-flux} (b) is nearly constant at $\Pi(k,t)\simeq 11\pm 1$ and is consistent with the energy dissipation rate $\varepsilon$ in the inertial range. These results indicate that the incompressible kinetic energy steadily flows in wave number space through the Richardson cascade at the constant rate $\Pi$, and finally dissipates to compressible excitations at the rate $\varepsilon\simeq \Pi$. This energy flow is shown in the diagram of Fig. \ref{fig-energy-current}. The energy flux $\Pi$ is slightly smaller than the energy dissipation rate $\varepsilon$, which may be attributable to the weak interaction between vortices and sound waves.

\begin{figure}[htb] \centering \begin{minipage}[t]{0.5\linewidth} \begin{center} \includegraphics[width=.99\linewidth]{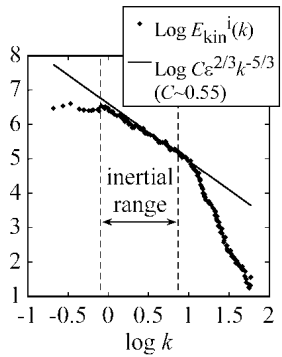} \end{center} \end{minipage}
\caption{\label{fig-steady-Kolmogorov} Energy spectrum $E\sub{kin}\up{i}(k,t)$ for quantum turbulence. The plotted points are from an ensemble average of 50 randomly selected states at $t>25$. The solid line is the Kolmogorov law $C\varepsilon^{2/3}k^{-5/3}$.} \end{figure}

To complete this picture of QT, we calculated the energy spectrum $E\sub{kin}\up{i}(k)$. The result is plotted in Fig. \ref{fig-steady-Kolmogorov}. Because the case is similar to that of decaying turbulence, the energy spectrum is quantitatively consistent with the Kolmogorov law in the inertial range $2\pi/X_0<k<2\pi/\xi$, which is equivalent to $0.79\lesssim k\lesssim 6.3$.
This inertial range is slightly narrower than that of the decaying turbulence because the moving random potential sets the energy-containing range. The resulting Kolmogorov constant $C$ is $C=0.55\pm 0.07$, which is smaller than that in CT as well as that for decaying turbulence. These smaller Kolmogorov constants $C$ in both decaying and steady superfluid turbulence than that in CT may be characteristic of QT. We propose clarifying that as one of our future works.

\section{Conclusion}\label{sec-conclusion}

In this paper, we argue that QT can be a prototype of CT. To support this argument, we used a modified GP equation to investigate the dynamics of QT and compared the results to those of CT. In the original GP equation, turbulence is suppressed by compressible short-wavelength excitations. To overcome this problem, we introduced a dissipation term that has effects only at scales smaller than the healing length and then numerically studied the modified GP equation. The main results are as follows.
\begin{enumerate}[(i)]
  \item The dissipation term does not disturb the dynamics of quantized vortices and enables one to investigate the statistics of quantized vortices in QT free from short-wavelength excitations.
  \item For decaying turbulence that starts from a random phase configuration of the wave function, the dissipation is effective for compressible excitations. This results in a incompressible kinetic energy spectrum that is consistent with the Kolmogorov law.
  \item A steady QT can be made by introducing energy injection as a moving random potential. In both steady and decaying turbulence, the incompressible kinetic energy dominates the total kinetic energy because of the dissipation term. The energy flux of the incompressible kinetic energy has a constant value in the inertial range and is consistent with the energy dissipation rate at small scales. The energy spectrum of the incompressible kinetic energy is also quantitatively consistent with the Kolmogorov law in the inertial range. These two results support the picture of QT shown in Fig. \ref{fig-energy-current}.
\end{enumerate}
These results indicate that QT is similar to CT and thus QT can become a prototype of turbulence for understanding the inertial range, the Kolmogorov law, and the Richardson cascade process. We hope that these results will lead to new insights into the nature of turbulence.

\section*{Acknowledgment}

We acknowledge W. F. Vinen and Toshiyuki Gotoh for useful discussions. MT and MK acknowledge the support of research grants from the Japan Society for the Promotion of Science (Grants No. 15340122 and 1505983, respectively).

\end{document}